\documentstyle[12pt]{article}

\newcommand{\be}{\begin{equation}}
\newcommand{\ee}{\end{equation}}
\newcommand{\bea}{\begin{eqnarray}}
\newcommand{\eea}{\end{eqnarray}}
\newcommand{\sptwo}{1.4}
\newcommand{\doublespace}{\edef\baselinestretch{\sptwo}\Large\normalsize}
\newcommand{\newsection}[1]{
\section{#1}
\setcounter{equation}{0}}
\renewcommand{\theequation}{\thesection.\arabic{equation}}

\newcounter{newapp}
\renewcommand{\thenewapp}{\Alph{newapp}}

\begin{document}
\hspace*{\fill} PURD-TH-93-14 \\
\hspace*{\fill} VAND-TH-93-14 \\
\hspace*{\fill} November 1993 \\
\begin{center}
{\large\bf Strongly Interacting Longitudinal Gauge Bosons\\
 In \\
Supersymmetric Models}
\end{center}
{}~\\
\begin{center}
{\bf T.E. Clark}\\
{\it Department of Physics\\
Purdue University\\
West Lafayette, IN 47907-1396}
{}~\\
{}~\\
{and}
{}~\\
{}~\\
{\bf W.T.A. ter Veldhuis}\\
{\it Department of Physics and Astronomy\\
Vanderbilt University\\
Nashville, TN 37235}
{}~\\
\end{center}
\begin{center}
{\bf Abstract}
\end{center}

A non-linear sigma model effective lagrangian is analyzed for theories in
which supersymmetry is softly broken at scales below the electroweak symmetry
breaking scale.  Besides the gauge and matter supermultiplets, the low energy
theory contains only three Goldstone chiral multiplets. The higgsino,
gaugino as well as the charged and neutral Higgs bosons have (light)
phenomenologically acceptable masses, the values of which depend on the
explicit soft supersymmetry breaking parameters.  In addition,
the longitudinal vector bosons become strongly interacting at high energies
($M_Z \ll E \ll 4\pi v$). The equivalence theorem is exploited in order to
obtain their scattering amplitudes.  Furthermore,
supersymmetry results in enhanced longitudinal vector boson production of
Higgs bosons.
\pagebreak

\doublespace

\newsection{Introduction}

The minimal supersymmetric standard model (MSSM) \cite{Nil}\cite{Hab}\cite{Bar}
compactly solves the
naturalness as well as the technical fine-tuning problems.  The electroweak
symmetry breaking is catalyzed by the soft (no quadratic divergence)
supersymmetry (SUSY) breaking terms arising from the hidden supergravity
sector.  As a consequence, the Higgs sector self-couplings are given by the
electroweak gauge coupling constants.  Hence, the model remains perturbative
up to the Planck scale.  Generalizations of the MSSM, whether motivated from
string inspired grand unified theories or from attempts to solve the
$\mu-$problem, involve additional electroweak matter multiplets.  In
particular, the new multiplets mix with the two Higgs superfields of the
MSSM so as to alter the mass spectrum.  The paradigm for such a process is
the minimal plus singlet supersymmetric standard model, (M+1)SSM
\cite{Der}\cite{Drees}.
The tree
level upper bound of the lightest neutral scalar is raised from $M_Z$ to a
bound that depends on the singlet-doublet Higgs interaction strength, $g$, and
the ratio of Higgs doublet vacuum expectation values, $\tan\beta \equiv
{v_B\over v_T}$,
$$M_h^2  =  M_Z^2 \left(\cos^2 {2\beta} +{2g^2\over g_1^2 +g_2^2}\sin^2
{2\beta} \right).$$
In this case, as in the MSSM , the internal symmetries are to remain
perturbatively natural up to the GUT scale.  The coupling constant $g$
has an infrared quasi-fixed point value of approximately 0.87 \cite{Ellis},
which leads
to a lightest Higgs mass upper bound of around 120 GeV
\cite{Bin}\cite{Ellw}\cite{TtV}\cite{Elliott}\cite{Mor}\cite{Esp}.

More complex extensions of the MSSM follow patterns similar to that of the
(M+1)SSM \cite{Esp2}\cite{Kane}.  Detailed parameterizations of the mass
spectrum and the soft SUSY
breaking parameter values have been addressed in numerous studies.  In all
cases the parameter space is investigated that provides a SUSY breaking
scale above that of the electroweak scale and, more specifically, a
perturbative mechanism for electroweak symmetry breaking that remains such
up to GUT scales.  In the MSSM there is no alternative, it is a prescribed
part of the model.  In the (M+1)SSM, it is the raison d'etre for the SUSY
model, but as such, a matter of choice.  In this paper we desire to study
the situation in which SUSY remains unbroken at energies below the
electroweak symmetry breaking scale, $\Lambda =4\pi v$, with
$v\approx 250$ GeV.  Since, in the broken symmetry phase, only the
partners of the electroweak Goldstone bosons need be light, we
will be considering the heavy mass (triviality) limit for the remaining
particles.
The low energy effective theory consequently contains fewer SUSY multiplets of
particles than the perturbative MSSM.  In
the (M+1)SSM, this means that  the entire singlet as well as the heavy
neutral Higgs, the pseudoscalar and their fermion partner become more
massive than $\Lambda$.  Without soft SUSY breaking the partners to the
Goldstone bosons, that is the remaining light neutral Higgs and the two
charged Higgs particles, are degenerate in mass with the $Z^0$ and the
$W^\pm$, respectively, as are their fermionic partners according to the
SUSY Higgs mechanism.  With soft SUSY breaking, the light neutral Higgs
field can acquire a mass higher than $M_Z$ and the charged Higgs fields
a mass higher than $M_W$.  The fermion partners along with the corresponding
gauginos,
however, will acquire masses half of which are lower and half of which are
higher than that of the gauge
bosons.

Of course, in lowest order, the low energy dynamics described above is
independent of the
particular short distance physics that gives rise to it, be it a strongly
interacting (M+1)SSM or some complicated dynamical symmetry breaking scheme,
as long as SUSY is softly broken at scales $M_{SUSY}\leq\Lambda$.  Hence,
the corresponding effective (softly broken) supersymmetric action is the
same, regardless of the mechanism for electroweak symmetry breaking.
Thus the characteristic mass spectrum for every such SUSY model will have,
besides massless Goldstone bosons, neutral and charged Higgs particles with
masses below
$\Lambda =4\pi v$.  In addition there will be
neutral and charged fermion partners with some masses in the $M_Z$ or $M_W$ to
$\Lambda$ range and an equal number with masses in the $0$ to $M_Z$ or $M_W$
range,
respectively.  The exact values,
as will be seen, depend on the explicit values of the
SUSY breaking parameters.

In the non-supersymmetric case, recall that the heavy Higgs limit of the
standard model can be described by the strong Higgs self-coupling limit of
the scalar fields.  Indeed, the scalar sector of the standard model can be
parameterized by the unitary $2\times 2$ matrix
\bea
U &\equiv & \pmatrix{ & \cr
\Big(\tilde\phi\Big) & \Big(\phi\Big)\cr
 & \cr}\nonumber\\
 & & \nonumber\\
 &=& \pmatrix{
h^{0\dagger} & h^+ \cr
-h^- & h^{0} \cr}
\eea
with
\be
\phi=\pmatrix{h^+\cr
h^0\cr}
\ee
the usual Higgs doublet field and
\be
\tilde\phi= \pmatrix{h^{0
\dagger}\cr
-h^-\cr}
\ee
its hypercharge conjugate doublet.  The scalar sector of the
standard model has the gauge invariant Lagrangian
\be
{\cal L}=\frac{1}{4}{\rm Tr}\left[D_\mu
U^\dagger D^\mu U\right] -\frac{\lambda}{4} \left[
\frac{1}{2}{\rm
Tr} \left[ U^\dagger U\right] -v^2\right]^2,
\ee
with the gauge covariant derivative
\be
D_\mu U=\partial_\mu U +ig_2\vec W_\mu\cdot \vec T  U-ig_1 B_\mu U T^3,
\ee
where $\vec T =\frac{1}{2}\vec\sigma$ and $\vec\sigma$ are the Pauli matrices.
The strong coupling limit,
$\lambda  =\frac{1}{2}(\Lambda/v)^2  \gg 1$, leads to
the constraint on the scalar
fields $\frac{1}{2}{\rm Tr} [U^\dagger U] =v^2$ with the
interpretation that only the Goldstone bosons remain in the theory at
energies below the electroweak symmetry breaking scale $\Lambda =4\pi v$ and
the gauge symmetry transformations are realized non-linearly.  Radiative
corrections to this tree effective lagrangian can be included by considering
${\rm Tr}\left[D_\mu U^\dagger D^\mu U\right]  $ to be the lowest order term
in a derivative expansion of the effective lagrangian for momenta below
$\Lambda$.  In references \cite{Appel}-\cite{Long2}
detailed lists of all dimension four through
six operators are given, including the lowest order custodial $SU(2)$ symmetry
violating term ${\rm Tr}\left[ T^3 D_\mu U^\dagger D^\mu U\right]
$ whose coefficient describes the contributions of the gauge radiative
corrections  to $\Delta \rho$, as well as the new physics above
$\Lambda$.

The resulting derivatively coupled non-linear $\sigma$-model effective
lagrangian describes strongly self-interacting Goldstone bosons at high
energy.  According to the equivalence
theorem \cite{Lee}\cite{Corn}\cite{Vay}\cite{Chan}, the scattering
amplitudes
involving longitudinally polarized gauge bosons are equivalent, up to
corrections of $O(\frac{M_W}{E})$, to the amplitudes with the longitudinal
gauge bosons, $W^\pm_L,~Z_L$ replaced by the corresponding Goldstone bosons,
$w^\pm,~z$, at high energies
\be
T(W_L,\dots,Z_L,\ldots) =T(w,\ldots,z,\ldots) +O(\frac{M_W}{E}).
\ee
Hence, these tree longitudinal gauge boson scattering amplitudes at high
energy can be read off directly from the non-linear Goldstone boson
lagrangian.  In the notation of an invariant length interval in the
symmetric space $SU(2)\times SU(2)/SU(2)$
\cite{boul}\cite{Ban}, the lagrangian is
\be
{\cal L}=\frac{1}{2} \partial_\mu \pi^i g_{ij}(\pi)\partial^\mu \pi^j,
\ee
with the choice of coordinates $U=\sigma {\bf 1} +2i\vec T\cdot\vec \pi$
and with the constraint $\frac{1}{2}{\rm Tr}[U^\dagger U ]= \det{U} =v^2$
yielding $\sigma =\sqrt{v^2-\vec\pi^2}$ (i.e. $h^0=\sigma -
i\pi^3 $ and $h^+= i(\pi^1-i\pi^2)$) .  This yields
the metric for this parametrization
of $U$: $g_{ij}=\delta_{ij} + \pi^i \pi^j /(v^2-\vec\pi^2)$.
When expanded in a power series in $\pi^i/v$, the enhanced longitudinal
gauge boson scattering amplitudes at high energy, $E \gg M_W$,  are simply
obtained \cite{Chan}\cite{Chan2}
\bea
T(W^+_LW^-_L\longrightarrow  Z_L Z_L) &\simeq & i\frac{s}{v^2}\nonumber\\
T(W^+_LW^-_L\longrightarrow W^+_LW^-_L) &\simeq & -i\frac{u}{v^2},
\eea
while
\be
T(Z_LZ_L\longrightarrow Z_LZ_L) \simeq 0.
\ee
The remaining longitudinal gauge boson amplitudes can be obtained by
crossing symmetry,  for example
\be
T(W^\pm_LZ_L\longrightarrow W^\pm_LZ_L) \simeq i \frac{t}{v^2}.
\ee
Hence, we are led to the optimistic alternatives of a light (perturbative)
Higgs boson ($M_h<\Lambda$) being observed directly, or, if heavy ($M_h>
\Lambda$), the enhanced scattering of longitudinal gauge bosons.

Analogously, the K\"ahler potential describing the non-linear realization of
the supersymmetric $SU(2)\times U(1)$  electroweak gauge symmetry can be
written in terms of the two Higgs doublet chiral superfields (recall that
$\sigma$ and $\vec\pi$ are chiral superfields here)
\bea
H_B &=& \pmatrix{H^+\cr
H^0_B\cr}\nonumber\\
 &=& \pmatrix{ i\pi^+\cr
\sigma-i\pi^3\cr}
\eea
and
\bea
H_T &=& \pmatrix{H^0_T\cr
H^-\cr}\nonumber\\
 &=& \pmatrix{\sigma+i\pi^3\cr
i\pi^-\cr},
\eea
grouped to form the matrix chiral superfield $U$
\bea
U &=&\sigma {\bf 1} + 2i\vec T\cdot \vec \pi \nonumber\\
 &=& \pmatrix{ & \cr
\big(H_T\big) &\big(H_B\big)\cr
 & \cr}\nonumber\\
 &=&\pmatrix{H^0_T & H^+\cr
H^- & H^0_B\cr} =\pmatrix{\sigma +i\pi^3 & i(\pi^1 -i\pi^2 )\cr
i(\pi^1 +i\pi^2 ) & \sigma -i\pi^3\cr}.
\eea
The matrix is constrained,
\be
\det {U}= H_T \epsilon
H_B = \sigma^2 + \vec\pi^2 =\frac{1}{2}v_T v_B=\frac{1}{4}v^2
\sin{2\beta},
\ee
so that a non-linear realization of $SU(2)\times U(1)$ is induced on $\vec\pi$
\be
U^\prime =LUR^{-1},
\ee
with $L=e^{i\vec T\cdot \vec \Lambda}$ and $R=e^{iT^3\Lambda_Y}$,
where the chiral superfields $\vec\Lambda\,(\Lambda_Y)$ parameterize the
$SU(2)\,(U(1))$ gauge transformations.  The gauge invariant K\"ahler
potential is made from powers
of the two independent $SU(2)\times U(1)$ gauge invariant
superfields \cite{Fer}
\bea
X &=& {\rm Tr}\left[ \bar Ue^{-2g_2\vec T\cdot \vec W}\, U\, e^{2g_1 T^3 Y}
\right]\nonumber\\
Y &=& {\rm Tr}\left[ T^3 \,\bar Ue^{-2g_2\vec T\cdot \vec W}\, U\, e^{2g_1
T^3 Y}\right],
\eea
where $\vec W$ and $Y$ are
the $SU(2)$ and $U(1)$ gauge fields, respectively.  Hence
the most general K\"ahler potential is given by
\be
K=\sum_{m,n=0}^\infty K_{mn} X^mY^n,
\ee
which yields the lowest order terms in a derivative expansion of the action,
$\Gamma =\int dV K$.  Note that $\Gamma =\int dV\, X$ is the simplest such
action.  As well it has the form of the MSSM Higgs fields' kinetic energy
and so preserves $\rho =1$ at the tree level.  Indeed, $Y$ will involve
violations of $\rho=1$, as can be seen most easily in the unitary gauge
(here $v_T=v_B$ and SUSY is unbroken in order of simplify, inessentially,
the algebra)
\bea
X\vert_{Unitary} &=& v^2\left[1+g_2^2W^+W^- +\frac{1}{2}(g_2^2 +g_1^{2})Z^2
\right]\nonumber\\
Y\vert_{Unitary} &=& -v^2 \sqrt{g_2^2 +g_1^{2}}Z.
\eea
So, for instance, the $Y^2$ term in $K$ yields a non-trivial $\Delta \rho$:
$\Delta\rho \sim K_{02}$.

In section 2 we will
describe the non-linear sigma model effective action for supersymmetric
electroweak symmetry breaking.  The mass spectrum in the SUSY Higgs and gaugino
sector is fully determined in terms of the soft SUSY breaking parameters
and $M_Z$ and $M_W$.  At the tree level the gauge field and (s)matter sector
masses are as in the MSSM.  In section 3 the equivalence
theorem relating the scattering amplitudes for longitudinal vector bosons
to the corresponding Goldstone boson amplitudes is exploited.  Although the
presence of light Higgs fields is possible, the longitudinal Goldstone
bosons are still strongly
interacting \cite{Rui}.  The tree amplitudes for longitudinal
$W^\pm$ and $Z^0$ scattering are shown to grow with energy:
\bea
T(W^+_LW^-_L\longrightarrow  Z_L Z_L) &\simeq & i \sin^2 {2\beta}\,\,
\frac{s}{v^2}\nonumber\\
T(W^+_LW^-_L\longrightarrow W^+_LW^-_L) &\simeq & -i \sin^2 {2\beta}\,\,
\frac{u}{v^2},
\eea
while
\be
T(Z_LZ_L\longrightarrow Z_LZ_L) \simeq 0.
\ee
Since SUSY is broken at comparable scales, it is found that the scattering
of longitudinal gauge bosons into neutral and charged Higgs bosons is
similarly enhanced.  In fact, the scattering amplitude for longitudinal
$Z^0$ bosons to produce light Higgs bosons, $Z_L\, Z_L\longrightarrow h\, h$,
is the dominant mode to neutrals since, for $s>M_h^2$, the amplitude for
$Z_L\, Z_L \longrightarrow Z_L\, Z_L$ vanishes, as noted, while the Higgs
production amplitudes grow with $s$
\bea
T(Z_LZ_L\longrightarrow hh) &\simeq & -i\sin^2 {2\beta}\,\,\frac{2s}{v^2}
\nonumber\\
T(W^+_LW^-_L\longrightarrow hh)&\simeq & -i\sin^2 {2\beta}\,\,
\frac{s}{v^2}.
\eea

The appendices recall some facts about non-linear realizations of gauge
symmetries in supersymmetric theories.
In appendix A  the heavy Higgs limit of the
(M+1)SSM is shown to lead to the supersymmetric
non-linear sigma model with a chiral Goldstone
superfield for each broken generator.
The explicit form of the component field non-linear sigma model
lagrangian is derived in appendix B.
The Killing vectors in different coordinate systems for the
$SU(2)\times U(1)/U(1)$ K\"ahler manifold are discussed in
appendix C.
\newpage

\newsection{The Effective Action }

The K\"ahler manifold describing the non-linear realization of
spontaneously broken electroweak symmetry breaking in supersymmetric
theories has the chiral superfield coordinates $\pi^i\,,\,i=1,2,3$.  (We
will deal with the
completely doubled realization  \cite{Buc} of the electroweak gauge
symmetry; see reference \cite{Khl} for a discussion of the possibility of a
minimal realization with non-trivial
fixed points.)  As suggested in the introduction, a specific
coordinate system can be chosen by taking the heavy Higgs limit of the
(M+1)SSM (see appendix A), although all the low energy physics is
re-parameterization invariant and so it does not depend on this specific
choice.  Hence,  we introduce the Goldstone chiral supermultiplets
\bea
H_B & \equiv & \pmatrix{H^+ \cr H^0_B\cr} =\pmatrix{i\pi^+ \cr
\sigma- i\pi^0\cr}\nonumber\\
H_T & \equiv & \pmatrix{H^0_T \cr H^-\cr} =\pmatrix{\sigma + i\pi^0
\cr  i\pi^- \cr},
\label{HTHB}
\eea
with
\bea
\pi^\pm & = &\pi^1 \mp i\pi^2 \nonumber\\
 & & \nonumber\\
\sigma & = &\sqrt{\frac{1}{2}v_Tv_B -\vec \pi^2} = v\sqrt{\frac{1}{4}
\sin{2\beta} -\vec\pi^2/v^2},
\label{sigma}
\eea
where the vacuum values of the supermultiplets are given by
\bea
<H_B> & = & \frac{1}{\sqrt{2}}\pmatrix{0 \cr v_B\cr}\nonumber\\
<H_T> & =  & \frac{1}{\sqrt{2}}\pmatrix{v_T \cr 0 \cr}
\eea
and the ratio of vacuum values defines the angle $\beta$ through
$\tan{\beta} =v_B/v_T$ and the electroweak
vacuum value $v^2 =v_T^2 +v_B^2$.  The
electroweak gauge transformations have
the usual form for the chiral superfields
\bea
H_B^\prime & = & e^{\frac{i}{2}\Lambda_Y}e^{i\vec\Lambda\cdot\vec T}H_B
\nonumber\\
H_T^\prime & = & e^{-\frac{i}{2}\Lambda_Y}e^{i\vec\Lambda\cdot\vec T}H_T .
\eea
Due to the above constraint,
$H_T \epsilon H_B =\sigma^2 +\vec\pi^2 =\frac{1}{2}v_Tv_B$, this
is actually a non-linear realization
on the Goldstone superfields $\pi^i$.  The $SU(2)$
gauge fields, $W^i\,,i=1,2,3$, and the $U(1)$ gauge field ,$Y$, transform as
\bea
e^{-2g_2\vec W^\prime\cdot\vec T} & = & e^{i\vec{\bar\Lambda}\cdot\vec T}
e^{-2g_2\vec W\cdot\vec T}e^{-i\vec\Lambda\cdot\vec T}\nonumber\\
e^{-g_1Y^\prime} & = & e^{\frac{i}{2}\bar\Lambda_Y}e^{-g_1Y}e^{-\frac{i}{2}
\Lambda_Y}.
\eea
The remaining quark and lepton matter superfields transform as in the MSSM, see
appendix B.  Defining the combination of electroweak gauge fields
\bea
V_B & =& \frac{1}{2}g_1Y +g_2\vec T\cdot\vec W\nonumber\\
V_T & =& -\frac{1}{2}g_1Y +g_2\vec T\cdot\vec W,
\eea
the fundamental gauge invariant
Goldstone superfield terms are given by $\left(\bar H_B
e^{-2V_B}H_B\right)$ and $\left(\bar H_Te^{-2V_T} H_T\right)$.  The remaining
matter field kinetic energy and Yukawa
terms are as in the MSSM and so are relegated
to Appendix B.

The simplest lowest order Goldstone superfield effective action, $\Gamma =
\int dV K$, is given by the K\"ahler potential
\be
K=\bar H_Te^{-2V_T} H_T + \bar H_Be^{-2V_B}H_B .
\ee
The most general soft SUSY
breaking terms \cite{Gir} (soft in that only logarithmic corrections
in the  momentum $\Lambda$ occur) are
the $\rho =1$ preserving, $\theta\,\bar\theta$
independent terms from $K$ and the
$\theta\,\bar\theta$ independent, $\Delta\rho$
producing terms from $\left(\bar H_Te^{-2V_T} H_T - \bar H_Be^{-2V_B}H_B
\right)$.
Hence the gauge invariant Goldstone superfield action can be written as
\bea
\Gamma_G &=& \int dV \,  [1+a\theta^2\bar\theta^2]K\nonumber\\
 &   &+ \int dV \, b\theta^2\bar\theta^2 \left[\bar H_Te^{-2V_T} H_T - \bar H_B
e^{-2V_B}H_B\right],
\label{Gamma_G}
\eea
with $a$ and $b$ the only SUSY breaking
parameters in the pure Goldstone sector.
When the gaugino soft SUSY breaking mass
terms and the Yang-Mills auxiliary field terms are included,
the vacuum and the Goldstone bosons',
electroweak gauge bosons' and their superpartners' mass spectra
can be determined.  In components the superfield K\"ahler
potential action, including the gaugino mass
lagrangian, ${\cal L}_{YM\rlap /S}$, and Yang-Mills
auxiliary $D_T$ field and $D_B$ field lagrangian
terms,  ${\cal L}_{YMD}$, reduces to the
usual K\"ahler form of the Higgs SUSY Lagrangian (see Appendix B):
\be
\Gamma_K =\int d^4x{\cal L}_K
\label{action1}
\ee
where
\be
{\cal L}_K ={\cal L}_{GS}
+{\cal L}_{G\rlap /S} + {\cal L}_{YMD} + {\cal L}_{YM\rlap /S}
\label{action2}
\ee
with
\be
\Gamma_G =\int dx({\cal L}_{GS} +{\cal L}_{G\rlap /S})
\label{action3}
\ee
and
\bea
{\cal L}_{GS} & =  &
\left(D_\lambda A_T \right)^\dagger\left(D^\lambda A_T \right)+
i\overline{\psi_T}\overline{\rlap /D} \psi_T +F_T^\dagger F_T\nonumber\\
 & & -A_T^\dagger D_T A_T +\sqrt{2}\left[\overline{\psi_T}\bar\lambda_T A_T +
A_T^\dagger \lambda_T \psi_T\right]
+\left( T\longrightarrow B\right).
\label{action4}
\eea
The generic chiral superfield in components is given by
\be
H=e^{-i\theta\rlap /\partial \bar\theta}
\left[A-i\sqrt{2}\theta^\alpha\psi_\alpha +
\theta^2 F\right],
\label{comp}
\ee
where the complex first component has the general real pseudoscalar field, $P$,
and real scalar field,
$S$, structure $A=P-iS$.  The soft SUSY breaking terms for the
Goldstone multiplets from equation (\ref{Gamma_G})
define ${\cal L}_{G\rlap /S}$, and are simply
given by the component field lagrangian
\be
{\cal L}_{G\rlap /S}=(a+b)A_T^\dagger A_T +(a-b)A_B^\dagger A_B.
\label{action5}
\ee
The auxiliary  gauge field and the gaugino soft SUSY breaking terms from the
Yang-Mills sector are also considered in the
calculation of the vacuum state and
Goldstone and gauge supermultiplet masses.
The corresponding Lagrangian for the
auxiliary fields is
\be
{\cal L}_{YMD} = -\frac{1}{2}\vec D_W\cdot\vec D_W-\frac{1}{2}D^2_Y,
\label{action6}
\ee
while the gaugino mass SUSY breaking terms are
\bea
{\cal L}_{YM\rlap /S } & = &
\frac{1}{2}\tilde m_W \vec\lambda_W\cdot\vec\lambda_W
+\frac{1}{2}\tilde m_Y\lambda^2_Y\,+\,h.c. \nonumber\\
 & = & \frac{1}{2}\pmatrix{\lambda_\gamma & \lambda_Z\cr}
\pmatrix{\tilde m_\gamma & \tilde m_{\gamma Z}\cr \tilde m_{\gamma Z} & \tilde
m_Z\cr}\pmatrix{\lambda_\gamma \cr \lambda_Z\cr}
+ \tilde m_W \lambda_+ \lambda_- +\,h.c. .
\label{action7}
\eea
The mass and electroweak gaugino eigenfields are related, as usual, via the
weak mixing angle $\theta_W$ with
$\tan{\theta_W}=g_1/g_2$ and charged fields $\lambda_\pm
=\frac{1}{\sqrt{2}}(\lambda_1 \mp i\lambda_2 )$
\bea
\lambda_\gamma &=& \lambda_3\sin{\theta_W}+\lambda_Y\cos{\theta_W}\nonumber\\
\lambda_Z &=& \lambda_3\cos{\theta_W}-\lambda_Y\sin{\theta_W},
\eea
so that the soft SUSY breaking gaugino masses are related by
\bea
\tilde m_\gamma &=&
\left(\tilde m_W\sin^2{\theta_W}+\tilde m_Y\cos^2{\theta_W}\right)\nonumber\\
\tilde m_Z &=& \left(\tilde m_W\cos^2{\theta_W}+\tilde m_Y
\sin^2{\theta_W}\right)\nonumber\\
\tilde m_{\gamma Z} &=&
\frac{1}{2}\left(\tilde m_W -\tilde m_Y\right)\sin{2\theta_W}.
\eea

The masses of the Goldstone-Higgs boson and gaugino-higgsino sectors can be
determined from above.
Eliminating the constrained fields and shifting the Higgs
fields by their vacuum expectation values,
the kinetic energy terms for the neutral
Higgs and higgsino fields acquire a finite wavefunction renormalization factor
(see Appendix B).  Rescaling the $\psi_3$ field by this factor,
\be
\psi_3 \longrightarrow \frac{1}{2}\sqrt{1+\sin{2\beta}}\,\,\, \psi_3 ,
\ee
the neutral fermion mass matrix in the
$(\lambda_Z ,\,\psi_3 ,\,\lambda_\gamma )$ basis becomes
\be
\tilde M =\pmatrix{\tilde m_Z & M_Z & \tilde m_{\gamma Z}\cr
M_Z &0 & 0\cr
\tilde m_{\gamma Z} & 0 & \tilde m_\gamma\cr},
\ee
while the charged fermion mass matrix
in the complex basis $(\lambda_+ ,\,\psi_+ )$ is
\be
\tilde M_{ch}=\pmatrix{\tilde m_W & M_W \sqrt{2}\cos{\beta}\cr
M_W \sqrt{2}\sin{\beta} & 0\cr}.
\ee
The gauge boson masses are found to be  given by their usual form
\bea
M_W &=& \frac{1}{2}g_2\,v\nonumber\\
M_Z &=& \frac{1}{2} \sqrt{g_1^2 +g_2^2}\,v.
\eea

The neutralino matrix is Hermitian and
can be diagonalized directly.  For simplicity we choose
$\tilde m_{\gamma Z}=0$, implying that $\tilde m_W =\tilde m_Y$ and hence
$\tilde m_Z =\tilde m_W=\tilde m_\gamma $,
the photino mass, $\tilde M_\gamma$, is simply given
by the breaking term $\tilde M_\gamma = \tilde m_\gamma$.
The squares of the zino mass,
$\tilde M_Z$, and the higgsino mass, $\tilde M_h$ are
\bea
\tilde M_Z^2 & = & M_Z^2\left[1+\frac{1}{2}\frac{\tilde
m_Z^2}{M_Z^2}\left(\sqrt{1+4\frac{M_Z^2}{\tilde m_Z^2}}+1\right)\right]
\nonumber\\
\tilde M_h^2 & = & M_Z^2\left[1-\frac{1}{2}\frac{\tilde
m_Z^2}{M_Z^2}\left(\sqrt{1+4\frac{M_Z^2}{\tilde m_Z^2}}-1\right)\right] .
\eea
The product of these neutralino masses yields the relation
\be
\tilde M_h \tilde M_Z =M_Z^2 .
\ee
As seen above, $\tilde M_Z \geq M_Z $
while $\tilde M_h \leq M_Z$.  This places an upper bound
phenomenological restriction on the photino mass,
$\tilde M_\gamma$, since $\tilde M_h$ cannot be too
small.  These masses are plotted in Figure 1 as a function of the photino mass.
\begin{figure}
   \vspace{2.5in}
   \caption{The neutralino masses as a function of the photino mass.}
\end{figure}

Multiplying the chargino mass matrix by its Hermitian conjugate,
$\tilde\mu_{ch}^2 = \tilde M_{ch}^\dagger \tilde M_{ch}$,
the squared mass matrix
is determined
\be
\tilde\mu^2_{ch} =
\pmatrix{\tilde m_W^2 +2M_W^2 \sin^2\beta & \sqrt{2} \tilde m_W M_W
\cos\beta\cr
\sqrt{2} \tilde m_W M_W \cos\beta & 2M_W^2 \cos^2\beta \cr}.
\ee
The charged fermion squared mass eigenvalues are found to be
\bea
\tilde M^2_{W^\pm} &=& M_W^2 \left[ 1 + \frac{1}{2}\frac{\tilde
m_W^2}{M_W^2}\left(\sqrt{1+4\frac{M_W^2}{\tilde m_W^2}+4\frac{M_W^4}{\tilde
m_W^4}\cos^2{2\beta}} + 1\right)\right]\nonumber\\
\tilde M^2_{h^\pm} &=& M_W^2 \left[ 1 - \frac{1}{2}\frac{\tilde
m_W^2}{M_W^2}\left(\sqrt{1+4\frac{M_W^2}{\tilde m_W^2}+4\frac{M_W^4}{\tilde
m_W^4}\cos^2{2\beta}} - 1\right)\right].
\eea
The product of these chargino masses yields the relation
\be
\tilde M_{W^\pm} \tilde M_{h^\pm} = M_W^2 \sin{2\beta} .
\label{inoprod}
\ee
Thus, one chargino mass is greater
than $M_W$ while the other is less than $M_W$.
For small $\sin{2\beta}$
(i.e., $\beta$ close to $0$ or $\frac{\pi}{2}$), the charged higgsino mass
becomes unacceptably small.
This results in phenomenological restrictions on the values of $\beta$.  The
chargino masses are plotted in
Figure 2 as a function of the photino mass for various values of $\beta$.

\begin{figure}
   \vspace{2.5in}
   \caption{The chargino masses as a function of the photino mass
 for various values of $\beta$.}
\end{figure}

The Goldstone scalar fields' effective potential is given by
\bea
V_G &=& -\frac{1}{2}D_Y^2
-\frac{1}{2}\vec D_W \cdot \vec D_W -D^A J_A\nonumber\\
 & & +(a+b)A_T^\dagger A_T +(a-b)A_B^\dagger A_B ,
\eea
where we have included the
soft SUSY breaking terms $a$ and $b$ and have trivially eliminated the
auxiliary fields, $F^i$,
by means of their equations of motion.  Recall that the first component of the
Goldstone field gauge current \cite{Cla} is given by equation (\ref{J_A})
\be
J_A =A_T^\dagger T^A_T A_T +A_B^\dagger T^A_B A_B .
\ee
{}From Appendix B, this becomes
\bea
V_G &=& \frac{1}{2}g_2^2
\left( A_T^\dagger T^i A_T -A_B^\dagger T^i A_B \right)^2 +\frac{1}{8}
g_1^2  \left( A_T^\dagger A_T - A_B^\dagger A_B \right)^2\nonumber\\
 & & \qquad +(a+b)A_T^\dagger A_T +(a-b)A_B^\dagger A_B \nonumber\\
 &=& \frac{(g_1^2 +g_2^2 )}{8}
\left( A_T^\dagger A_T -A_B^\dagger A_B \right)^2 +\frac{g_2^2}{2}
\left( A_T^\dagger A_B \right)^2 \nonumber\\
 & & \qquad +(a+b)A_T^\dagger A_T +(a-b)A_B^\dagger A_B .
\eea
Expanding about the vacuum values
\bea
<H_T> &=& \frac{1}{\sqrt{2}}\pmatrix{v_T  \cr
0}\nonumber\\
<H_B> &=& \frac{1}{\sqrt{2}}\pmatrix{0 \cr
v_B},
\eea
or in terms of the doublet and SUSY component fields' vacuum values
\bea
<A_\sigma > &=& \sigma \nonumber\\
<A^3> &=& -i \eta ,
\eea
where the vacuum values $\sigma$ and $\eta$ are related to $v_T$ and $v_B$ by
\bea
\frac{1}{\sqrt{2}}v_T  &=&
\sigma + \eta = \frac{1}{\sqrt{2}}v \cos{\beta}\nonumber\\
\frac{1}{\sqrt{2}} v_B &=& \sigma - \eta = \frac{1}{\sqrt{2}} v \sin{\beta},
\eea
the minimum of the effective potential is found to occur at
\be
-\frac{1}{2}M_Z^2 = a + b\sec{2\beta},
\label{min}
\ee
which relates the soft SUSY
breaking parameters $a$ and $b$ to the $Z$ mass and the angle $\beta$.
Note that $b=0$ for $\beta=\frac{\pi}{4}$ ($v_T =v_B$),
and is non-zero for unequal values of $v_T$
and $v_B$, that is when $<A^3> =-i\eta \neq 0$.

The scalar field mass matrix
decouples into a charged scalar
field and a neutral scalar field matrix.  In the
complex charged electroweak basis,
$A_-^\dagger =(A^{1\dagger} -iA^{2\dagger}),\, A_+ =(A^{1}-
iA^{2})$, the Hermitian mass matrix is
\be
M_{ch}^2 = \left( M_W^2 +2a \right)
\pmatrix{{\sin}^2{\beta} & -\frac{1}{2}\sin{2\beta} \cr
-\frac{1}{2}\sin{2\beta} & {\cos}^2{\beta} \cr}.
\ee
This can be diagonalized by introducing the Goldstone boson fields
\bea
w_+ &\equiv & \left( A_-^\dagger \cos{\beta} +A_+ \sin{\beta}\right)
\nonumber\\
w_- &\equiv & \left(A_- \cos{\beta} +A_+^\dagger \sin{\beta}\right) ,
\label{w+-}
\eea
which are massless
(in the Stueckelberg gauge or have mass $M_W$ in the Feynman-${\rm R}_\xi$
gauge) and the charged Higgs
fields (SUSY partners to the charged Goldstone bosons $w_\pm$)
\bea
h_+ &\equiv & \left( A_+ \cos{\beta} -A_-^\dagger \sin{\beta}\right)\nonumber\\
h_- &\equiv & \left( A_+^\dagger \cos{\beta} -A_- \sin{\beta}\right) ,
\label{h+-}
\eea
with mass squared
\be
M_{h^\pm}^2 = M_W^2 +2a .
\ee
Note that $M_{h^\pm}^2 \ge 0$
requires that $a \ge -\frac{1}{2}M_W^2$.  Further, the
phenomenological lower
bound for $M_{h^\pm}$ implies a lower bound for $a$.  The charged Higgs
mass is plotted in Figure 3
as a function of $\sqrt{|a|}$ for positive and negative values of $a$.

\begin{figure}
   \vspace{2.5in}
   \caption{The charged
Higgs mass as a function of the soft SUSY breaking parameter,
$(\pm)|a|^{\frac{1}{2}}$,
the $+$ denoting positive values of $a$ and the $-$ denoting
negative values of $a$.}
\end{figure}

These fields can be grouped
to make manifest the rotational nature of their linear combinations
\be
\pmatrix{w_- \cr h_- \cr}
=\pmatrix{\cos\beta & \sin\beta\cr-\sin\beta & \cos\beta\cr}\pmatrix{A_- \cr
A_+^\dagger\cr},
\ee
and its Hermitian conjugate relation
\be
\pmatrix{w_+ & h_+ \cr} =
\pmatrix{A_-^\dagger & A_+ \cr}\pmatrix{\cos\beta & -\sin\beta\cr
\sin\beta & \cos\beta\cr}.
\ee

\begin{figure}
   \vspace{2.5in}
   \caption{The neutral Higgs mass as a function of the charged Higgs mass
 for various values of $\beta$.}
\end{figure}

The neutral scalar field mass matrix in the $A_3 ,\, A_3^\dagger$ basis is
\be
\frac{1}{2} M^2 =
\frac{\left(M_Z^2 -2a{\sin}^2{2\beta}\right)}{1+\sin{2\beta}}\pmatrix{1 & -1\cr
-1 & 1 \cr}.
\ee
Recall that we must rescale
the neutral fields by a finite wavefunction renormalization factor.  This
follows from the form of the K\"ahler metric in the vacuum
\bea
{g_{\bar i j}}\vert_{A_\sigma =\sigma ,\, A^3 =h} &=& 2\delta_{ij}
+\frac{2\eta^2}{\sigma^2}\delta_i^3 \delta_j^3\nonumber\\
 &=& \left\{  \begin{array}{cc}
 2\delta_{ij} & \mbox{for $ i,j =1,2$ } \\
\frac{4}{1+\sin{2\beta}} &  \mbox{for $i=j=3$} \\
0 & \mbox{otherwise}
\end{array}
\right. .
\eea
Introducing the renormalized neutral Goldstone boson field
\be
z \equiv \sqrt{\frac{2}{1+\sin{2\beta}}}\left( A^3 +A^{3\dagger}\right)
\ee
and the renormalized neutral Higgs field (SUSY partner to the $z$)
\be
h \equiv i \sqrt{\frac{2}{1+\sin{2\beta}} }\left( A^3 -A^{3\dagger}\right),
\ee
the matrix is diagonalized
with $z$ massless (in the Stueckelberg gauge or having mass $M_Z$ in the
Feynman-${\rm R}_\xi$ gauge) and $h$ having mass squared
\be
M_h^2 =M_Z^2 +2a{\sin}^2{2\beta} .
\label{hmass2}
\ee
Note that $M_h^2 \ge 0$
requires that $a {\sin}^2{2\beta} \ge -\frac{1}{2}M_Z^2 $,
which is always satisfied for $a \ge -\frac{1}{2}M_W^2$.
Using these masses,
the minimum condition for the potential, equation (\ref{min}), can
be written as
\be
-\frac{1}{2} \left( M_Z^2 - M_W^2 + M_{h^\pm}^2 \right) =b\sec{2\beta},
\ee
while the neutral Higgs mass squared, equation (\ref{hmass2}), is given by
\be
M_h^2 = M_Z^2 + \left( M_{h^\pm}^2 - M_W^2 \right){\sin}^2{2\beta} .
\label{hmass22}
\ee
Note that if $M_h \leq M_Z$,
then $M_{h^\pm} \leq M_W$ and vice versa.  Finally, using equation
(\ref{inoprod}) for the chargino masses,
the neutral Higgs mass equation (\ref{hmass22}) becomes
\be
\left( M_h^2 -M_Z^2 \right)M_W^4  =
\left( M_{h^\pm}^2 -M_W^2 \right)\tilde M^2_{W^\pm} \tilde
M^2_{h^\pm} .
\ee
The neutral Higgs mass is plotted in
Figure 4 as a function of the charged Higgs mass for various values
of $\beta$.
\newpage

\newsection{Longitudinal Vector Boson Scattering}

The equivalence theorem
provides a simple means to calculate the high energy scattering amplitudes for
the longitudinal gauge bosons
$W_L^\pm ,\, Z_L$.  They are equal to the same amplitudes with the
longitudinal vector bosons
replaced by their corresponding Goldstone fields, $W_L^\pm \rightarrow
w_\pm ,\, Z_L \rightarrow z$.
Since the Goldstone bosons are derivatively coupled to each other, the
dominant contributions to
their high energy, $E \gg M_Z$, scattering amplitudes can be obtained from the
kinetic energy terms in their K\"ahler lagrangian
\be
{\cal L} =\partial_\lambda A^{i\dagger}
g_{\bar i j}(\vec A) \partial^\lambda A^j ,
\ee
with the metric
\bea
g_{\bar i j} &=& 2\left\{ \delta_{ij}
+{\partial A^\dagger_\sigma \over \partial A^{i\dagger}}{\partial
A_\sigma \over \partial A^j}\right\}\nonumber\\
 &=& 2\left\{ \delta{ij}
+ {A^{i\dagger} A^j \over A^\dagger_\sigma A_\sigma}\right\}\nonumber\\
 &=& 2\left\{\delta_{ij}
+ {A^{i\dagger} A^j \over \sqrt{\frac{1}{2}v_Tv_B -\vec{\bar A}^2}
\sqrt{\frac{1}{2}v_Tv_B -\vec A^2}} \right\}.
\eea

In the tree approximation,
the dominant contributions arise from direct quartic field terms and from one
particle exchange graphs made
of trilinear field terms.  Hence, the metric must be expanded about the
vacuum values, recall
$<A^3> =-i\eta =\frac{-i}{\sqrt{8}}(v_T -v_B)$, through second order in the
fields.  Writing the complex
fields $A^i$ in terms of their shifted real components $P^i ,\, S^i$ as
\be
A^i =P^i -iS^i -i\eta\delta^i_3 ,
\ee
we find through bilinear in fields
\bea
g_{\bar i j} &=& \delta_{ij} +\frac{\eta^2}{\sigma^2}\delta^3_i\delta^3_j -
\frac{2\eta^3}{\sigma^4}S^3
\delta^3_i\delta^3_j +\frac{i\eta}{\sigma^2} \left[ \delta^3_i \left(P^j -i
S^j\right)-\delta^3_j\left( P^i +i S^i\right)\right] \nonumber\\
 & & +\frac{1}{\sigma^2}
\left[ P^i P^j +S^i S^j +i\left( S^iP^j -P^i S^j \right)\right] \nonumber\\
 & &  -\frac{2i\eta^2}{\sigma^4}
\left[ \delta^3_i \left( P^j -iS^j\right) -\delta^3_j \left( P^i
+iS^i\right)\right]S^3\nonumber\\
 & & +\frac{\eta^2}{\sigma^4}
\delta^3_i \delta^3_j \left[ P_1^2 +P_2^2 -S_1^2 -S_2^2 +\left(1-
\frac{2\eta^2}{\sigma^2}\right) P^2_3 -
\left(1-\frac{4\eta^2}{\sigma^2}\right) S^2_3 \right] .\nonumber\\
 & &
\label{metric2}
\eea
Recall $\eta=\frac{1}{\sqrt{8}}(v_T -v_B )
=\frac{v}{\sqrt{8}}( \cos\beta -\sin\beta )$ and $\sigma
=\frac{1}{\sqrt{8}}(v_T +v_B )=
\frac{v}{\sqrt{8}}( \cos\beta +\sin\beta )$ so that
\bea
8\eta^2 &=& v^2 \left( 1-\sin{2\beta}\right)\nonumber\\
8\sigma^2 &=& v^2 \left( 1+\sin{2\beta}\right) .
\eea
The charged Goldstone bosons,
$w_\pm$, and the physical charged Higgs bosons, $h_\pm$, are given by
linear combinations of the complex
$A^i$ fields as in equations (\ref{w+-}) and (\ref{h+-}).  Hence, the
charged component fields $A_\pm =P_\pm -iS_\pm$, that is
\bea
P_- &=& \frac{1}{2}
\left( A_- +A_+^\dagger \right)  ,\,\,\,\, S_- = \frac{i}{2} \left( A_- -
A^\dagger_+\right)\nonumber\\
P_+ &=&  \frac{1}{2}\left( A_-^\dagger +A_+
\right)  ,\,\,\,\, S_+ = \frac{i}{2} \left( A_+ -A^\dagger_-
\right),
\eea
are related to the mass eigenstate fields by
\bea
P_\pm &=& \frac{1}{2} \left( \cos\beta
+\sin\beta \right) w_\pm +\frac{1}{2} \left( \cos\beta -\sin\beta
\right) h_\pm \nonumber\\
S_\pm &=& \mp \frac{i}{2}
\left( \cos\beta -\sin\beta \right) w_\pm \pm\frac{i}{2} \left( \cos\beta
+\sin\beta \right) h_\pm .
\eea
Moreover, in order to cast the kinetic terms of the
neutral Goldstone boson and Higgs boson in their
conventional form a finite wavefunction renormaliztion
is required:
\bea
P^3 &=& \frac{1}{2}\sqrt{\frac{1+\sin{2\beta}}{2}} z\nonumber\\
S^3 &=&  \frac{1}{2} \sqrt{\frac{1+\sin{2\beta}}{2}} h .
\eea
After some algebra,
the Feynman rules can be gleaned from the Lagrangian.
The various scattering amplitudes  are calculated
in the Stueckelberg gauge in which the Goldstone bosons are massless
\bea
T(w^+w^- \rightarrow w^+w^-) & = &
\frac{i}{v^2} \frac{2\sin^2 2\beta}{1+\sin 2\beta}
\left[ s+t - \cos^2(\frac{\pi}{4}+\beta)
\left\{\frac{s^2}{s-M_h^2} \right.\right. \nonumber\\
 & & \left.\left.
\qquad\qquad\qquad\qquad\qquad\qquad +\frac{t^2}{t-M_h^2}\right\}
\right], \nonumber\\
T(zz\rightarrow w^+w^-) & = &
\frac{i}{v^2} \frac{2\sin^2 2\beta}{1+\sin 2\beta}
\left[ s- \cos^2(\frac{\pi}{4}+\beta)
\left\{\frac{s^2}{s-M_h^2}\right\}
\right], \nonumber\\
T(zz\rightarrow zz) & = &
\frac{i}{v^2} \frac{2\sin^2 2\beta}{1+\sin 2\beta}
\left[ s+t+u - \cos^2(\frac{\pi}{4}+\beta)
\left\{\frac{s^2}{s-M_h^2} \right.\right.\nonumber\\
 & &  \left.\left. \qquad\qquad\qquad\qquad\qquad +\frac{t^2}{t-M_h^2}
+\frac{u^2}{u-M_h^2}\right\}
\right]. \nonumber\\
 & &
\label{scatt1}
\eea

Besides the enhancement of
the pure Goldstone boson scattering processes at high energy, SUSY implies
a similar enhancement for the
Goldstone boson to Higgs boson production processes.  The tree
approximation leading contributions
also follow from the second order expansion of the metric, equation
(\ref{metric2}).  After some algebra,
the Higgs production amplitudes are secured
\bea
T(zz\rightarrow h^0h^0)  & = &
-\frac{i}{v^2} \frac{2\sin^2 2\beta}{1+\sin 2\beta}
\left[ 2s -
\cos^2(\frac{\pi}{4}+\beta)
\left\{s\frac{s-2M_h^2}{s-M_h^2}
 \right.\right. \nonumber \\
 &  & \left.\left. \qquad\qquad
-\frac{(t-M_h^2)^2}{t}-\frac{(u-M_h^2)^2}{u}+6\frac{M_h^4}{\sin2\beta}\right\}
\right], \nonumber \\ & & \nonumber \\
T(zz\rightarrow h^+h^-)  & = &
-\frac{i}{v^2} \frac{2\sin^2 2\beta}{1+\sin 2\beta}
\left[ s -
\cos^2(\frac{\pi}{4}+\beta)
\left\{ \frac{s^2}{s-M_h^2}\right\}
\right], \nonumber \\  & & \nonumber \\
T(w^+w^-\rightarrow h^0h^0)  & = &
-\frac{i}{v^2} \frac{2\sin^2 2\beta}{1+\sin 2\beta}
\left[ s -
\cos^2(\frac{\pi}{4}+\beta)
\left\{ s\frac{s-2M_h^2}{s-M_h^2}\right\}
\right] \nonumber \\
 &   &
-\frac{i}{v^2} \frac{2\cos^2 2\beta}{1+\sin 2\beta}
\left[
\cos^2(\frac{\pi}{4}+\beta)
\left\{ \frac{M_h^4}{u-M_{h^\pm}^2} +
\frac{M_h^4}{t-M_{h^\pm}^2} \right\}
\right], \nonumber \\  & & \nonumber \\
T(w^+w^-\rightarrow h^+h^-)  & = &
-\frac{i}{v^2} \frac{2\sin^2 2\beta}{1+\sin 2\beta}
\left[ s -
\cos^2(\frac{\pi}{4}+\beta)
\left\{ \frac{s^2}{s-M_h^2}\right\}
\right] \nonumber \\
 &   &
+\frac{i}{v^2} \frac{2}{1+\sin 2\beta}
\left[ t -
\cos^2(\frac{\pi}{4}+\beta)
\left\{ t\right\}
\right]  \nonumber \\
 &   &
+\frac{i}{v^2} \frac{2\cos^2 2\beta}{1+\sin 2\beta}
\left[ t -
\cos^2(\frac{\pi}{4}+\beta)
\left\{ \frac{t^2}{t-M_h^2}\right\}
\right]\nonumber\\
T(w^\pm w^\pm \rightarrow h^\pm h^\pm )  &
= & +\frac{4i}{v^2} \frac{\cos^2 2\beta}{1+\sin 2\beta}
\left[ t+u- \cos^2(\frac{\pi}{4}+\beta)
\left\{\frac{t^2}{t-m_0^2}+\frac{u^2}{u-m_0^2}\right\}
\right] \nonumber \\
 &  &
+\frac{2is}{v^2} \nonumber\\
T(z w^\pm \rightarrow h h^\pm )  & = &  -\frac{t}{v^2} \sin{2\beta}
-\frac{m_0^2}{v^2} \frac{1-\sin{2\beta}}{1+\sin{2\beta}}
+\frac{2 m_0^2}{v^2} \frac{\sin{2\beta}}{1+\sin 2\beta}
\cos^2(\frac{\pi}{4}+\beta) .\nonumber\\
 & &
\label{scatt2}
\eea

For energies above the
charged and neutral Higgs masses, as well as $M_Z$, equation (\ref{scatt1})
reduces to the scattering
amplitudes for strongly interacting longitudinal gauge bosons
\bea
T(W_L^+W_L^-\rightarrow W_L^+W_L^-) & \approx &
-\frac{iu}{v^2} \sin^2 2\beta , \nonumber \\
T(Z_LZ_L \rightarrow W_L^+W_L^-) & \approx &
+\frac{is}{v^2} \sin^2 2\beta , \nonumber \\
T(Z_LZ_L \rightarrow Z_LZ_L) & \approx & 0.
\eea
Furthermore, the enhanced  Higgs
boson production amplitudes from longitudinal vector boson scattering,
equation (\ref{scatt2}), become
\bea
T(Z_LZ_L \rightarrow h^0h^0) & \approx &
-\frac{2is}{v^2} \sin^2 2\beta , \nonumber \\
T(Z_LZ_L \rightarrow h^+h^-) & \approx &
-\frac{is}{v^2} \sin^2 2\beta , \nonumber \\
T(W_L^+W_L^- \rightarrow h^0h^0) & \approx &
-\frac{is}{v^2} \sin^2 2\beta , \nonumber \\
T(W_L^+W_L^- \rightarrow h^+h^-) & \approx &
-\frac{is}{v^2} \sin^2 2\beta
+\frac{it}{v^2} (1+\cos^2 2\beta)\nonumber\\
T(W_L^\pm W_L^\pm \rightarrow h^\pm h^\pm )
& \approx & +\frac{2is}{v^2} \sin^2 2\beta
\nonumber\\
T(Z_L W_L^\pm \rightarrow h h^\pm )  & \approx & -\frac{t}{v^2} \sin 2\beta .
\eea
\smallskip

The authors wish to thank
A. Papadopoulos for numerous interesting as well as useful conversations.
This work was supported in part by the U.S. Department
of Energy under contracts DE-AC02-76ER01428 (Task B) and DE-FG05-8SER40226.
\newpage

\setcounter{equation}{0}
\renewcommand{\theequation}{\thenewapp.\arabic{equation}}

\section*{Appendix A \,\,\,  The Heavy Higgs Limit}

The SUSY non-linear realization of the electroweak symmetry breakdown,
$SU(2) \times U(1) \rightarrow U(1)$, can be obtained directly by
considering the heavy Higgs limit of the (M+1)SSM. The Higgs sector is
comprised of the two $SU(2)$ doublet chiral superfields, with opposite
weak hypercharge, $H_B$ and $H_T$, and an
additional singlet chiral superfield, N.
The supersymmetric action for these fields is given by
\bea
\Gamma & = & \int dV \left[ \bar{H}_B H_B + \bar{H}_T H_T + \bar{N} N \right]
\nonumber \\
 & - & \int dS \left[  \xi N + \frac{1}{2} m N^2 + \frac{\lambda}
{3!}N^3 + \mu H_T\epsilon H_B - g N H_T\epsilon H_B \right] +h.c. \nonumber\\
 & &
\label{action}
\eea
The superfield Euler-Lagrange equations are given by field
derivatives of the superpotential so that
\bea
\bar{D}\bar{D} \bar{N} & = & \xi +mN + \frac{\lambda}{2} N^2 -g H_T\epsilon
H_B,
\nonumber \\
\bar{D}\bar{D} \bar{H}_B & = & (\mu -gN) H_T \epsilon , \nonumber \\
\bar{D}\bar{D} \bar{H}_T & = & (\mu -gN)\epsilon H_B.
\eea
In the strong coupling limit, the superpotential derivative terms on
the right hand sides above must vanish for a non-vanishing effective action.
Hence we find the contraints
\bea
N & = & \frac{mu}{g}\equiv n < \infty, \nonumber \\
H_T\epsilon H_B & = & \frac{\xi}{g} + \frac{m \mu}{g^2} + \frac{\lambda}{2}
\frac{\mu^2}{g g^2} \equiv \frac{1}{4} v^2 \sin{2 \beta} < \infty.
\label{constraint}
\eea
The heavy singlet field becomes stationary at its vacuum value
in the strong coupling limit $\mu,g \rightarrow \infty$ with
$n=\frac{mu}{g}$ fixed. Likewise the chiral composite field
$H_T\epsilon H_B$ becomes
stationary at its vacuum value as $g \rightarrow \infty$
with at least one or all $\frac{\xi}{g}$, $\frac{m}{g}$, $\frac{\lambda}{g}$
fixed. This results in the heavy neutral scalar field, the pseudoscalar
field and their heavy neutral fermion partner field in
$H_B$ and $H_T$ to become heavy. This can be made clearer by introducing
the $\sigma$-model notation of equation (\ref{HTHB}). Hence
$H_T\epsilon H_B = \sigma^2 + \vec{\pi}^2=\frac{1}{2} v_T v_B$, allowing the
$\sigma$ superfield to be eliminated as in equation (\ref{sigma}),
$\sigma = \sqrt{\frac{1}{2} v_B v_T - \vec{\pi}^2}$, from the action,
inducing non-linear $\pi^i$ self interactions from the $\sigma$-kinetic
energy terms, as is familiar.

Recalling the component field expansion for $\sigma$ and $\vec{\pi}$,
\bea
\sigma & = & e^{-i \theta \rlap /\partial \bar{\theta}}
\left[ A_{\sigma} -i\sqrt{2}\theta \psi_{\sigma} + \theta^2 F_{\sigma}
\right], \nonumber \\
\pi^i & = & e^{-i \theta \rlap /\partial \bar{\theta}}
\left[ A^i -i\sqrt{2}\theta \psi^i+ \theta^2 F^i
\right].
\label{comp2}
\eea
The constraint equation
(\ref{constraint} ) can be expanded to yield the component
field constraints
\bea
A_{\sigma}^2 + \vec{A}^2 & = & \frac{1}{2} v_T v_B, \nonumber \\
A_{\sigma} \psi_{\sigma} + \vec{A}\cdot\vec{\psi} & = & 0, \nonumber \\
A_{\sigma} F_{\sigma} +\vec A\cdot\vec F
- \frac{1}{2}\psi_{\sigma}^2 -\frac{1}{2}
\vec{\psi}\cdot\vec{\psi} & = & 0.
\eea

The heavy Higgs limit can be further clarified by considering the
strong coupling limit of the component field
action. In particular the mass matrix will yield a heavy
mass for the singlet multiplet as well as for the $\sigma$-multiplet,
that is the heavier neutral scalar, the neutral pseudoscalar and
their neutral fermion partner field combinations of the $H_B$
and $H_T$ doublets. The linear $\sigma$-model action is given by
\bea
\Gamma & = & \int d^4x \left[
\partial_{\lambda} A_B^{\dagger} \partial^{\lambda} A_B +
\partial_{\lambda} A_T^{\dagger} \partial^{\lambda} A_T +
\partial_{\lambda} A_N^{\dagger} \partial^{\lambda} A_N + \right.
\nonumber \\
 & + & \frac{i}{2} \psi_B \stackrel{\leftrightarrow}{\rlap /\partial}
\bar{\psi}_B
+ \frac{i}{2} \psi_T \stackrel{\leftrightarrow}{\rlap /\partial}
\bar{\psi}_T
+ \frac{i}{2} \psi_N \stackrel{\leftrightarrow}{\rlap /\partial}
\bar{\psi}_N \nonumber \\
 & + & F_B^{\dagger} F_B + F_T^{\dagger} F_T + F_N^{\dagger} F_N \nonumber \\
 & + & \left\{ F_N
\left[ 4 g A_T\epsilon A_B -4\xi -4mA_N -2\lambda A_N^2 \right]
\right. \nonumber \\
 & + & 4 (g A_N - \mu)
\left[ F_T\epsilon  A_B +  A_T \epsilon F_B - \psi_T\epsilon \psi_B
\right] \nonumber \\
 & + & \left. \left.
2 (m + \lambda A_N)\psi_N^2
- 4 g \psi_N \left[ A_T\epsilon \psi_B +\psi_T \epsilon A_B
\right] +h.c. \right\} \right].
\eea
Note that the strong coupling limit
$g,\xi \rightarrow \infty$ with $\frac{\xi}{g}$ fixed yields the
doublet fields' constraint equations (\ref{constraint})
\bea
A_T\epsilon  A_B & = & \frac{\xi}{g} \equiv \frac{1}{2} v_T v_B \nonumber \\
A_T \epsilon \psi_B  + \psi_T \epsilon A_B & = & 0 \nonumber \\
F_T \epsilon A_B +  A_T \epsilon F_B - \psi_T \epsilon \psi_B & = & 0
\label{constraint2}
\eea
while $\lambda,m \rightarrow \infty$ and $\frac{2m}{\lambda}$
fixed yields the decoupled stationary massive
singlet multiplet $A_N=\frac{2m}{\lambda}\equiv n, \psi_N=0, F_N=0$.
The resulting SUSY non-linear
$\sigma$-model lagrangian is given by the kinetic energy terms of the
constrained, equation (\ref{constraint2}), doublet fields
\bea
{\cal L} &=& \partial_{\lambda} A_B^{\dagger} \partial^{\lambda} A_B +
\partial_{\lambda} A_T^{\dagger} \partial^{\lambda} A_T
 + \frac{i}{2} \psi_B \stackrel{\leftrightarrow}{\rlap /\partial}
\bar{\psi}_B
+ \frac{i}{2} \psi_T \stackrel{\leftrightarrow}{\rlap /\partial}
\bar{\psi}_T  \nonumber \\
 &  & + F_B^{\dagger} F_B + F_T^{\dagger} F_T  .
\eea
 \newpage

\setcounter{newapp}{2}
\setcounter{equation}{0}

\section*{Appendix B\,\,\,\,  The K\"{a}hler Potential}

The heavy Higgs limit of the (M+1)SSM will
give rise to the simplest form of the lowest order
SUSY non-linear sigma model action.
The (M+1)SSM action can be written as
\bea
\Gamma_{\rm (M+1)SSM} & = & \Gamma_G + \Gamma_M +
\Gamma_{YM} + \Gamma_{FI}
\eea
above $\Gamma_{YM}$, is the usual electroweak
gauge field kinetic energy terms and soft $SUSY$ breaking
gaugino mass terms, along with the supersymmetric
gauge fixing and Fadeev-Popov pieces.
$\Gamma_{FI}$ is a possible $U(1)$ hypercharge Fayet-Illiopoulous
term. $\Gamma_M$ is the gauge invariant supersymmetric quark and
lepton superfield kinetic energy terms as well as their
Yukawa-interactions along with their associated soft SUSY breaking terms.
Likewise $\Gamma_G$ are the Higgs doublet $H_B$, $H_T$ and
singlet $N$ gauge invariant supersymmetric kinetic
energy terms as well as the $H_B$, $H_T$, $N$ superpotential
terms and all the related soft SUSY breaking terms.
Hence the Higgs field
action terms are the gauge invariant version of those given in equation
(\ref{action})
\bea
\Gamma_G & = & \int dV \left[ Z_B \bar{H}_B e^{-2V_{H_B}}\bar{H}_B +
Z_T \bar{H}_T  e^{-2V_{H_T}}\bar{H}_T + Z_N \bar{N}N \right]\nonumber \\
 & + & \left\{ \int dS \left[ \xi N + \frac{1}{2}  m N^2 + \frac{1}{3!}
\lambda N^3 + \mu H_T\epsilon H_B
- g N H_T\epsilon H_B \right] + h.c. \right\}.\nonumber\\
 & &
\eea
The residual supergravity soft-SUSY breaking terms are
included as $\theta$, $\bar{\theta}$ dependent wavefunction
renormalization factors,
$Z_{\phi}=(1+m_{\phi}^2\theta^2\bar{\theta}^2)$, and
coupling constants, $g=g+a_g \theta^2$, $\mu=\mu+ a_{\mu}\theta^2$,
etc., mimicking the constant
graviton and gravitino coupling effects \cite{Gir}\cite{Bag}.  The matter
field terms can be
similarly written. They do not
involve the singlet field and are as in the MSSM:
\bea
\Gamma_M & = & \int dV \left[ Z_Q \bar{Q} e^{-2V_Q}Q +
Z_{T^C} \bar{T}^C e^{-2V_{T^C}}T^C +
Z_{B^C} \bar{B}^C e^{-2V_{B^C}}B^C \right. \nonumber \\
 & + &  \left. Z_{E^C} \bar{E}^C e^{-2V_{E^C}}E^C +
Z_L \bar{L} e^{-2V_L}L\right] \nonumber \\
 & + & \left\{ \int dS \left[ g_E^{mn} E_m^C L_n^a H_B^a +
g_B^{mn} B_m^C Q_n^a H_B^a
+ g_T^{mn} T_m^C Q_n^a H_{T}^a \right] + h.c. \right\}.\nonumber\\
 & &
\eea
where as before the wavefunction renormalization factors include the
soft SUSY breaking masses,
$Z_{\phi}=(1+m_{\phi}^2\theta^2\bar{\theta}^2)$, and the coupling
constants have the form $g=g+a_g \theta^2$. The vector
gauge fields are defined as
\bea
V_Q & = &-g_3 \vec{G}.\frac{\vec{\lambda}}{2} -
g_2 \vec{W}.\frac{\vec{\sigma}}{2} - \frac{1}{6} g_1 Y \nonumber \\
V_{T^C}  & = & + g_3 \vec{G}.\frac{\vec{\lambda}}{2}+
\frac{2}{3} g_1 Y \nonumber \\
V_{B^C}  & = & + g_3 \vec{G}.\frac{\vec{\lambda}}{2}-
\frac{1}{3} g_1 Y \nonumber \\
V_{E^C} & = & -g_1 Y \nonumber \\
V_L & = & -g_2 \vec{W}.\frac{\vec{\sigma}}{2} + \frac{1}{2} g_1 Y\nonumber\\
V_B &=& +\frac{1}{2}g_1 Y+g_2\vec W \cdot\vec T\nonumber\\
V_T &=& -\frac{1}{2}g_1 Y+g_2\vec W \cdot\vec T ,
\label{YM}
\eea
with the $U(1)$ generators
represented by the identity, the $SU(2)$ generators given by the Pauli
matrices, $T^i =\frac{\sigma^i}{2},\, i=1,2,3$,
and the $SU(3)$ generators represented by the Gell-Mann
matrices, $L^a =\frac{\lambda^a}{2},\, a=1,\ldots,8.$

The gauge invariant Yang-Mills terms have the structure
\bea
\Gamma_{YM}^{inv} & = & \int dS\, \frac{Z_W}{4g_2^2}\, {\rm Tr}\left[W^\alpha
W_\alpha \right] + \int dS\, \frac{Z_Y}{4g_1^2}\, Y^\alpha Y_\alpha + h.c. ,
\eea
with the field strenth spinors given as
\bea
W^\alpha & = & \bar{D}\bar{D}\left[e^{-2g_2\vec{W}.
\frac{\vec{\sigma}}{2}}D^\alpha e^{2g_2\vec{W}.\frac{\vec{\sigma}}{2}}
\right],
\eea
\bea
Y^\alpha & = & \bar{D}\bar{D}\left[e^{-2g_1 Y}D^\alpha
e^{2g_1 Y} \right] = 2 g_1 \bar{D}\bar{D}D^\alpha Y .
\eea
The gaugino mass terms
result from the soft SUSY breaking terms contained in the wavefunction
normalization factors
\bea
Z_W & = & \left(1+\frac{1}{2}\tilde{m}_W \theta\theta\right) ,
\nonumber \\
Z_Y & = & \left(1+\frac{1}{2}\tilde{m}_Y \theta\theta\right) .
\eea
The Fayet-Iliopoulous hypercharge term is simply
\bea
\Gamma_{FI} & = & \kappa \int dV Y.
\eea
Since, after eliminating
the auxiliary hypercharge field, $D_Y$, this is equivalent to the soft SUSY
breaking $b$-terms in the
Higgs action.  Hence, we set $\kappa =0$, incorporating the possibility of a
Fayet-Iliopoulous term
in the $b$-breaking terms. For a more detailed display of the MSSM and
(M+1)SSM notation conventions used here see \cite{Cla2}\cite{TtV}.

The strong coupling limit for the singlet
and doublet Higgs fields results in
the $\Gamma_{\rm (M+1)SSM}$ becoming a constrained field action.
Indeed, we have
that, in analogy to the discussion in Appendix A,
$\Gamma_{G}$ becomes simply the kinetic energy
terms for the constained doublet fields,
\bea
\Gamma_G & = & \int dV \left[
Z_{H_B} \bar{H}_B e^{-2V_{H_B}}H_B +
Z_{H_T} \bar{H}_T e^{-2V_{H_T}}H_T
\right]
\eea
where $H_T \epsilon H_B = \frac{1}{2} v_T v_B$ and we can write the
wavefunction renormalization factors as
$Z_{H_B}=1 + (a-b)\theta^2\bar{\theta}^2$,
$Z_{H_T}=1 + (a+b)\theta^2\bar{\theta}^2$.
The superpotential with the soft breaking terms and the singlet field
kinetic energy terms are a constant or zero. The remaining
action is as in the MSSM except for the
Yukawa interaction terms.  Since they involve the constrained
field, new interactions with the Goldstone superfields will occur as the
constrained fields are eliminated in favor of the Goldstone multiplet
fields.  The form of the (s)quark and (s)lepton mass terms will
however be the same as in the MSSM since the form of the
vacuum values of $H_B$ and $H_T$ are unchanged.

Generally the Goldstone multiplet action can be written in terms of
a gauge invariant
extension of the K\"ahler potential $K=K(\vec{\pi},\vec{\bar{\pi}})$
(see reference \cite{Cla} for notation conventions).
This can be accomplished most directly by simply implementing the
constraints
\bea
\sigma & = & \sqrt{\frac{1}{2}v_Tv_B-\vec{\pi}^2}, \nonumber \\
\bar{\sigma} & = & \sqrt{\frac{1}{2}v_Tv_B-\vec{\bar{\pi}}^2}.
\eea
Hence the gauge invariant extension of the super K\"ahler potential including
soft SUSY breaking terms is found
to be
\bea
K & = & Z_{H_B} \bar{H}_B e^{-2V_{H_B}}H_B +
Z_{H_T} \bar{H}_T e^{-2V_{H_T}}H_T,
\eea
with
\bea
H_B & = &
\left[
\begin{array}{c}
i\pi^+ \\
\sqrt{\frac{1}{2}v_Tv_B-\vec{\pi}^2}-i\pi^0
\end{array}
\right], \nonumber \\
H_T & = & \left[
\begin{array}{c}
\sqrt{\frac{1}{2}v_Tv_B-\vec{\pi}^2}+i\pi^0 \\
i\pi^-
\end{array}
\right].
\eea
The K\"ahler structure can be further manifested by expanding
the superfields in terms of their component fields as given in
equations (\ref{comp}).
The full action, $\Gamma_K$, (in the Wess-Zumino gauge)
including the Yang-Mills auxiliary field terms and the soft SUSY breaking
gaugino mass terms becomes (see equations (\ref{action1}-\ref{action4},
 \ref{action5}-\ref{action7}))
$\Gamma_{K} = \int d^4x {\cal L}_K$,
with ${\cal L}_K$ given by
\be
{\cal L}_K = {\cal L}_G +
{\cal L}_{YMD} + {\cal L}_{YM\rlap /S}
\ee
where $\Gamma_G =\int dx {\cal L}_G $ and
\be
{\cal L}_G = {\cal L}_{GS} + {\cal L}_{G\rlap /S}.
\ee
Expanding in terms of component fields, the individual component field
lagrangians are found to be
\bea
{\cal L}_{GS} & =  &
\left(D_\lambda A_T \right)^\dagger\left(D^\lambda A_T \right)+
i\overline{\psi_T}
\overline{\rlap /D} \psi_T +F_T^\dagger F_T  -A_T^\dagger D_T A_T \nonumber\\
 & &  +\sqrt{2}\left[\overline{\psi_T}\overline{\lambda_T} A_T +
A_T^\dagger \lambda_T\psi_T\right]
+\left( T\longrightarrow B\right)\nonumber\\
{\cal L}_{G\rlap /S} &=& (a+b)A_T^\dagger A_T +(a-b)A_B^\dagger A_B\nonumber\\
{\cal L}_{YMD} &=&
-\frac{1}{2}\vec D_W\cdot\vec D_W-\frac{1}{2}D^2_Y \nonumber\\
{\cal L}_{YM\rlap /S }
& = &\frac{1}{2}\tilde m_W \vec\lambda_W\cdot\vec\lambda_W +\frac{1}{2}\tilde
m_Y\lambda^2_Y\,+\,h.c. \nonumber\\
 & = & \frac{1}{2}\pmatrix{\lambda_\gamma & \lambda_Z\cr}
\pmatrix{\tilde m_\gamma & \tilde m_{\gamma Z}\cr \tilde m_{\gamma Z} & \tilde
m_Z\cr}\pmatrix{\lambda_\gamma \cr \lambda_Z\cr}
+ \tilde m_W \lambda_+ \lambda_- +\,h.c. ,
\nonumber\\
 & &
\label{lagr}
\eea
where the covariant derivatives follow from equation (\ref{YM})
\bea
D^\mu A_T &=& \partial^\mu A_T +iV_T^\mu A_T \nonumber\\
D^\mu \psi_T &=& \partial^\mu \psi_T +iV^\mu_T \psi_T ,
\eea
and analogously for $A_B$ and $\psi_B$.

The constraints can be solved to  yield the doublet component fields dependence
on $\vec A,\, \vec\psi,\, \vec F$.
This has the generic structure  (recall each chiral superfield, constrained
or not, has the component field
structure as defined in equation (\ref{comp},  \ref{comp2}))
\bea
A_T^a &=& H_T^a \vert_{\theta\bar\theta =0} \equiv H_T^a(\vec A)\nonumber\\
\psi_T^a &=& {\partial\over\partial\theta}
H_T^a\vert_{\theta\bar\theta =0}={\partial H_T^a(\vec A)\over
\partial A^i}\psi^i\nonumber\\
F_T^a &=& -\frac{1}{4}{\partial^2\over
\partial\theta^2}H_T^a\vert_{\theta\bar\theta =0}=
 \left\{F^i{\partial\over
\partial A^i}+\frac{1}{2}\psi^i\psi^j {\partial^2\over \partial A^i\partial
A^j}\right\} H^a_T (\vec A),
\eea
and similarly for $H_B$ and its
components.  In particular, we find that the constrained components of
$\sigma$  are given by
\bea
A_\sigma & = & \sqrt{\frac{1}{2} v_T v_B -\vec{A}^2} \nonumber \\
\psi_\sigma & = & - \frac{1}{\sqrt{\frac{1}{2} v_T v_B -\vec{A}^2}}
\vec{A}\cdot\vec{\psi} \nonumber \\
F_\sigma & = & - \frac{1}{\sqrt{\frac{1}{2} v_T v_B -\vec{A}^2}}
\left[ \vec{A}\cdot\vec{F} - \frac{1}{2}\vec{\psi}\cdot\vec{\psi}
-\frac{1}{2}\frac{(\vec{A}\cdot\vec{\psi})^2}
{\sqrt{\frac{1}{2} v_T v_B -\vec{A}^2}}
\right].
\eea
In addition, the $SU(2)\times U(1)$
gauge transformations of the doublet superfields will define the
Killing vector superfields,
$A_A^i(\vec\pi)$, in terms of the Goldstone superfields, $\pi^i$, (in this
notation we label $\Lambda_Y =\Lambda_4$,
so that the index $A$ runs over all the generator labels
1, 2, 3, 4 of $SU(2) \times U(1)$
while index $i$ runs over the broken generator labels 1, 2, 3 of
$SU(2) \times U(1)/ U(1)$ \cite{Cla})
\bea
\delta H_T  &=& (-\frac{i}{2}\Lambda_Y
+i\vec\Lambda\cdot\vec T)H_T=i\Lambda^A T^A_T
H_T\nonumber\\
 &\equiv & {\partial H_T\over
\partial \pi^i}\delta\pi^i = {\partial H_T\over \partial \pi^i}\Lambda^A
A_A^i(\vec\pi)\nonumber\\
\delta H_B &=&  (+\frac{i}{2}
\Lambda_Y +i\vec\Lambda\cdot\vec T)H_B=i\Lambda^A T^A_B
H_B\nonumber\\
 &\equiv & {\partial H_B
\over \partial \pi^i}\delta\pi^i = {\partial H_B\over \partial \pi^i}\Lambda^A
A_A^i(\vec\pi).
\label{delta}
\eea
Expanding in powers of
$\theta,\,\bar\theta$, the component field transformation equations are secured
\bea
iT^A_T A_T &=& {\partial H_T
(\vec A)\over \partial A^i}A_A^i (\vec A)\nonumber\\
iT^A_T \psi_T &=& \left({\partial H_T
(\vec A)\over \partial A^i}{\partial A^i_A(\vec A)\over \partial
A^j}+{\partial^2H_T(\vec A)\over
\partial A^j \partial A^i} A^i_A(\vec A)\right)\psi^j\nonumber\\
iT^A_T F_T &=& \left\{F^j{
\partial\over \partial A^j}+\frac{1}{2}\psi^j\psi^k {\partial^2\over \partial
A^j\partial A^k}\right\}
\left[{\partial H_T (\vec A)\over \partial A^i}A_A^i (\vec A)\right],
\eea
and likewise for the components of
$H_B$.  The specific form of the Killing vectors, $A^i_A$, will be
discussed in Appendix C.

These relations now allow
the component lagrangian ${\cal L}_K$ to be put into its manifest K\"ahler
form \cite{Wess}.  For example, the scalar field kinetic energy terms become
\bea
\partial_\lambda A_{T}^\dagger
\partial^\lambda A_{T} + \partial_\lambda A_{B}^\dagger
\partial^\lambda A_{B} &=&
\partial_\lambda A^{\dagger i}\left[{\partial \bar H_T^a\over \partial
A^{\dagger i}}{\partial H_T^a\over
\partial A^j}+ {\partial \bar H_B^a\over \partial A^{\dagger
i}}{\partial H_B^a\over \partial A^j}\right]\partial^\lambda A^j\nonumber\\
 &=&  \partial_\lambda A^{\dagger i} g_{\bar i j} \partial^\lambda A^j,
\eea
where the K\"ahler manifold
metric, $g_{\bar i j}$, is given, as above, by the $\theta\, \bar\theta$
independent component of the superfield K\"ahler metric
\bea
g_{\bar i j}(\vec{\bar\pi}, \vec\pi) &=& {\partial^2\over
\partial\bar\pi^i\partial\pi^j}K(\vec{\bar\pi},\vec\pi).
\eea
The simple super K\"ahler
potential (ignoring the SUSY breaking terms and the gauge couplings) is
$K(\vec{\bar\pi},\vec\pi)=
\bar H_T H_T +\bar H_B H_B = 2(\bar\sigma\sigma +\vec{\bar\pi}\cdot\vec\pi
)$, hence we find the component metric
\bea
g_{\bar i j} &=& 2\left(\delta_{ij}
+{\partial A_\sigma^\dagger\over \partial A^{\dagger i}}{\partial
A_\sigma\over \partial A^j}\right)\nonumber\\
 &=& 2\left( \delta_{ij}
+{A^{\dagger i}A^j \over \sqrt{\frac{1}{2}v_Tv_B -\vec{\bar
A}^2}\sqrt{\frac{1}{2}v_Tv_B -\vec{A}^2}} \right).
\eea
The gauge invariant scalar field kinetic
energy terms in equation (\ref{lagr}) yield those in terms  of the
complex Goldstone boson fields
\bea
{\cal L} &=& \partial_\lambda
A^\dagger_T\partial^\lambda A_T - i A^\dagger_T  V_T^\lambda
\partial_\lambda A_T  + i\partial_\lambda
A_T^\dagger  V_T^\lambda  A_T  \nonumber\\
 & & + A_T^\dagger V_{T\lambda} V_T^\lambda A_T + (T\rightarrow B)\nonumber\\
 &=& \left(D^\lambda A\right)^{\dagger i}g_{\bar i j}\left(D_\lambda
A\right)^j,
\eea
with the covariant derivative given by
\be
\left( D_\mu A\right)^i \equiv \partial_\mu A^i +V_\mu^A A_A^i.
\ee
The gauge fields have been
combined in the notation $V^A_\mu$ with (we choose $T^4_T =
-\frac{1}{2}$ and  $T^4_B = +\frac{1}{2}$ here)
\bea
V_T^\mu &=& -\frac{1}{2}g_1Y^\mu +g_2 \vec W^\mu\cdot \vec T\nonumber\\
 &\equiv & V^{A\mu} T^A_T\nonumber\\
V_B^\mu &=& +\frac{1}{2}g_1Y^\mu +g_2 \vec W^\mu\cdot \vec T\nonumber\\
 &\equiv & V^{A\mu} T^A_B,
\eea
so that the generalized vector field becomes
\be
V^A_\mu = \left\{ \begin{array}{cc}
g_2 W^i_\mu & \mbox{for $A=i=1,2,3$} \\
g_1 Y_\mu & \mbox{for $A=4$}
\end{array}
\right.
\ee
Analogously, each term in the
constrained field lagrangian can be expressed in terms of the component
fields.  The lagrangian ${\cal L}_K$ becomes
\bea
{\cal L}_K &=& \left(D^\lambda A
\right)^{\dagger i}g_{\bar i j}\left(D_\lambda A\right)^j +\frac{i}{2}
\bar\psi^i g_{\bar i j} \left( \overline{\rlap /D}\psi\right)^j\nonumber\\
 & &-\frac{i}{2}\left(
\overline{D^\mu \psi}\right)^i g_{\bar i j}\bar\sigma_\mu  \psi^j +D^A J_A
\nonumber\\
 & &+\sqrt{2}\left(
\bar\psi^i g_{\bar i j}A^j_A
\bar\lambda^A +\lambda^A\bar A^i_A g_{\bar i j} \psi^j
\right)+\frac{1}{4}R_{i\bar k j\bar l}
\left(\bar\psi^k\bar\psi^l\right)\left(\psi^i\psi^j\right)\nonumber\\
 & &+\left(\bar F^i +\frac{1}{2}\bar\Gamma^{\bar i}_{\bar m\bar n}\left(
\bar\psi^m\bar\psi^n\right)\right)
g_{\bar i j}\left(F^j
+\frac{1}{2}\Gamma^j_{rs}\left(\psi^r\psi^s\right)\right).
\eea
The fermion covariant derivative, $(D_\mu \psi)^i$, is defined to be
\be
(D_\mu \psi)^i = \partial_\mu
\psi^i +V^A_\mu {\partial A^i_A\over \partial A^j} \psi^j
+\Gamma^i_{jk}(D_\mu A)^j\psi^j .
\ee
The non-zero connection $\Gamma^i_{jk}$ is defined by
\bea
\Gamma_{\bar i jk} &=&
\left[ {\partial \bar H_T^a (\vec A^\dagger ) \over \partial A^{\dagger
i}}{\partial^2 H_T^a(\vec A)
\over \partial A^k \partial A^j} + {\partial \bar H_B^a (\vec A^\dagger )
\over \partial A^{\dagger i}}
{\partial^2 H_B^a(\vec A)\over \partial A^k \partial A^j}\right]\nonumber\\
 &=& {\partial g_{ik}\over \partial A^j}
= {\partial g_{ij}\over \partial A^k} = \Gamma_{\bar i kj},
\eea
with $\Gamma^i_{jk} = g^{\bar l i}
\Gamma_{\bar l jk}$, and correspondingly for the complex conjugate
connection $\bar\Gamma^{\bar i}_{\bar j \bar k}
=g^{\bar i l}{\partial g_{\bar k l}\over \partial
A^{\dagger j}}$.   The Riemann
curvature tensor, $R_{i\bar j k \bar l}$, for the K\"ahler manifold is
defined by $R_{i\bar j k \bar l}=g_{\bar l m}R^m_{i\bar j k}$ with
\be
R^m_{i\bar j k} ={\partial\over \partial A^{\dagger j}}\Gamma^m_{ik}.
\ee
Hence we find that
\be
R_{i\bar j k \bar l}
= {\partial\over \partial A^{\dagger j}}\Gamma_{\bar l ik}-\bar\Gamma^{\bar
n}_{\bar j \bar l}\Gamma_{\bar n ik}.
\ee
Finally the Killing potentials,
$J_A$, are given by the $\theta ,\, \bar\theta$ independent component of the
Goldstone superfield part
of the N\"oether gauge current  vector superfield \cite{Cla}
\bea
J_A (x,\theta,\bar\theta) &=& \bar H_T T^A_T H_T +\bar H_B T^A_B H_B\nonumber\\
 &=& \left\{-\frac{i}{2}A^i_A (\vec\pi) {\partial\over \partial \pi^i}
+\frac{i}{2}\bar A^i_A (\vec{\bar\pi})
{\partial\over \partial \bar\pi^i}\right\}\overbrace{\left[\bar H_T
H_T +\bar H_B H_B\right]}^{=K}  . \nonumber\\
 & &
\eea
Thus the Killing potentials become
\bea
J_A &=&J(x,0,0)=A^\dagger_T T^A_T A_T +A^\dagger_B T^A_B A_B\nonumber\\
 &=& -\frac{i}{2}\left( A^i_A{\partial
\over \partial A^i} -\bar A^i_A {\partial\over \partial A^{\dagger
i}}\right)\left[ A^\dagger_T A_T +A^\dagger_B A_B\right]\nonumber\\
 &=&  -i\left( A^i_A{\partial
\over \partial A^i} -\bar A^i_A {\partial\over \partial A^{\dagger
i}}\right)\left[ A^\dagger_\sigma A_\sigma
+\vec A^\dagger \cdot \vec A\right]\nonumber\\
 &=&  -i\left( A^i_A{\partial\over \partial A^i}
-\bar A^i_A {\partial\over \partial A^{\dagger
i}}\right)\left[ \sqrt{\frac{1}{2}v_Tv_B
-\vec{A}^{\dagger 2}}\sqrt{\frac{1}{2}v_T v_B -\vec A^2}
+\vec{A}^\dagger \cdot \vec A \right] \nonumber\\
 & = & i\left\{A^i\bar A^i_A -A^{i\dagger}
A^i_A +A^i A^i_A \frac{\sqrt{\frac{1}{2}v_T v_B -\vec
A^{\dagger 2}}}{\sqrt{\frac{1}{2}v_T v_B -\vec A^{2}}}
 -A^{i\dagger} \bar A^i_A
\frac{\sqrt{\frac{1}{2}v_T v_B -\vec A^{2}}} {\sqrt{\frac{1}{2}v_T v_B -
\vec A^{\dagger 2}}}\right\}.\nonumber\\
 & &
\label{J_A}
\eea
Differentiating the superfield
currents $J_A$ with respect to the Goldstone superfields, we find the
superfield differential equations
relating the Killing vectors and the Killing potentials (all superfields
here)
\bea
{\partial\over \partial\pi^i}J_A &=& i\bar A^j_A g_{\bar j i}\nonumber\\
{\partial\over \partial \bar\pi^i}J_A &=& -iA^j_A g_{\bar i j}.
\label{potential}
\eea

\newpage

\setcounter{newapp}{3}
\setcounter{equation}{0}

\section*{Appendix C\,\,\,\,  Coordinates And \\
\hspace*{2in}Killing Vectors}

The $SU(2)\times U(1)$ gauge
transformations of the $2\times 2$ matrix chiral superfield $U$,
\be
U^\prime = e^{i\vec T \cdot \vec\Lambda} U e^{-iT^3\Lambda_Y},
\ee
induce general coordinate
transformations of the Goldstone superfields used to parameterize the K\"ahler
manifold.  For the \lq\lq
sigma model" choice of coordinates in the body of the paper,
\be
U =\sigma {\bf 1} +2i\vec T \cdot \vec\pi ,
\label{U}
\ee
the infinitesimal gauge transformations yield
\bea
\delta\pi^1 &=&
+\Lambda^1 \sigma -\Lambda^2 \pi^3 +\Lambda^3 \pi^2 +\frac{1}{2}\Lambda_Y \pi^2
\nonumber\\
\delta\pi^2 &=&
+\Lambda^1 \pi^3 +\Lambda^2 \sigma -\Lambda^3 \pi^1 -\frac{1}{2}\Lambda_Y \pi^1
\nonumber\\
\delta\pi^3 &=&
-\Lambda^1 \pi^2 +\Lambda^2 \pi^1 +\Lambda^3 \sigma -\frac{1}{2}\Lambda_Y
\sigma  ,
\label{vari}
\eea
along with the constrained field's variation
\be
\delta\sigma = -\vec\Lambda \cdot \vec\pi +\frac{1}{2}\Lambda_Y \pi^3 .
\label{sigmavari}
\ee
(This also could be
obtained directly from the gauge transformations of $H_T$ and $H_B$, equation
(\ref{delta}).)
Indeed, since $\sigma$
is a constrained field, $\det{U} \equiv \frac{1}{2}v_T v_B $,  the
Goldstone multiplets
transform non-linearly upon substitution of $\sigma = \sqrt{\frac{1}{2}v_T v_B
-
\vec\pi^2}$ into equation
(\ref{vari}) and the $\sigma$ variation follows from their transformations
applied to the constraint equation (\ref{sigmavari}).

The chiral superfield Killing
vectors, $A_A^i (\vec\pi)$, define these non-linear transformations (here let
$\Lambda^4 =\frac{1}{2}\Lambda_Y$)
\be
\delta\pi^i = \Lambda^A \delta_A \pi^i \equiv \Lambda^A A_A^i (\vec\pi ).
\ee
Thus we secure
\be
A_A^i (\vec\pi)=\left\{ \begin{array}{ll}
-\epsilon^{iak} \pi^k +\delta^i_a \sigma & \mbox{for $A=a=1,2,3$} \\
-\epsilon^{i3k} \pi^k -\delta^i_3 \sigma & \mbox{for $A=4$}
\end{array}
\right. .
\ee
The Killing potential equation, (\ref{potential}), is satisfied by $A_A^i$.

The \lq\lq standard" coordinates
are defined by the exponential representation of the $2\times 2$ matrix
chiral superfield $U$ \cite{Fer}
\be
U\equiv fe^{2i\vec T \cdot \vec\xi}
=f\left[ \cos{\sqrt{{\vec\xi}^2}}+2i\vec T \cdot \vec\xi \,
\,{\sin{\sqrt{{\vec\xi}^2}}\over \sqrt{{\vec\xi}^2}} \right],
\ee
with
\be
\det{U} =f^2 =\frac{1}{2}v_T v_B
\ee
and $\xi^i$ the three Goldstone
boson superfields.  The transformation to the $\sigma$-model coordinates,
$\pi^i$, follows from the equality of the respective representations for $U$
\be
\pi^i = {\sin{\sqrt{{\vec\xi}^2}}\over \sqrt{{\vec\xi}^2}}\, f \xi^i .
\ee
The constraint equation
equality, $f \cos{\sqrt{{\vec\xi}^2}}=\sigma  =\sqrt{\frac{1}{2}v_T v_B -
{\vec\pi}^2}$, follows from
equation (\ref{U}).  Inverting the equation yields $\xi^i$ as a function of
$\pi^i$
\be
\xi^i = \rho (\vec\pi^2 ) \pi^i ,
\ee
with
\be
\rho (\vec\pi^2 ) = {\arcsin{
\sqrt{\frac{\vec\pi^2}{f^2}}}\over f\sqrt{\frac{\vec\pi^2}{f^2}}}.
\ee
{}From simple geometry,
note that $\sin{\sqrt{\vec\xi^2}}=\sqrt{\frac{\vec\pi^2}{f^2}}$, so that the
$\rho\sigma$ product is simply
\be
\rho\sigma = f\sqrt{\vec\xi^2}\cot{\sqrt{\vec\xi^2}}.
\label{rhosig}
\ee

The non-linear gauge transformations of the standard coordinates are
\bea
\delta\xi^i &=& {\partial\xi^i\over \partial\pi^j}
\delta\pi^j \equiv \Lambda^A P^i_j (\vec\xi )A_A^j
({\vec\xi}/{\rho})\nonumber\\
 &=& \left[\delta^{ij}
-\left(1-\frac{1}{\rho\sigma}\right)\frac{\xi^i \xi^j }{\vec\xi^2}\right]\rho
A_A^j ({\vec\xi}/{\rho}).
\eea
Exploiting the coordinate
transformations, $\xi^i =\rho\pi^i$, $\sigma=f\cos{\sqrt{\vec\xi^2}}$ and
equation (\ref{rhosig}),
the standard coordinate Killing vectors, $X_A^i (\vec\xi)$,  are secured
\be
\delta\xi^i \equiv \Lambda^A X_A^i (\vec\xi),
\ee
where $\rho = \sqrt{\vec\xi^2} \csc{\sqrt{\vec\xi^2}}$ and
\bea
X_A^i (\vec\xi) &=& P^i_j (\vec\xi )A_A^j ({\vec\xi}/{\rho})\nonumber\\
 &=&\left\{ \begin{array}{ll}
-\epsilon^{iak}\xi^k +\frac{\xi^i \xi^a}
{\vec\xi^2}+\left(\delta^i_a - \frac{\xi^i \xi^a}{\vec\xi^2}\right)
\sqrt{\vec\xi^2}\cot{\sqrt{\vec\xi^2}} & \mbox{for $A=a=1,2,3$}\\
-\epsilon^{i3k}\xi^k -\frac{\xi^i \xi^3}
{\vec\xi^2}-\left(\delta^i_3 - \frac{\xi^i \xi^3}{\vec\xi^2}\right)
\sqrt{\vec\xi^2}\cot{\sqrt{\vec\xi^2}} & \mbox{for $A=4$.}
\end{array}
\right. \nonumber\\
 & &
\eea
The simple K\"ahler potential can be written as
\be
K = {\rm Tr}{[\bar U U]}=
f^2 {\rm Tr}{\left[ e^{-i\vec{\bar\xi}\cdot \vec T}e^{i\vec\xi\cdot \vec
T}\right]}.
\ee
The K\"ahler manifold metric can be obtained from the potential
\be
g_{\bar i j} (\vec{\bar\xi}, \vec\xi )
= {\partial^2 K\over \partial \bar\xi^i \partial \xi^j} .
\ee

The gauge invariant potential is given by
\be
K= {\rm Tr}{\left[ \bar U
e^{-2g_2 \vec T \cdot \vec W} U e^{2g_1 T^3 Y}\right] }.
\ee
The superfield Killing potentials,
$J^A$, are the Noether gauge currents found from the auxiliary field
$D$-term coupling in $K$
\be
J^A = \left\{ \begin{array}{ll}
-2g_2{\rm Tr}{\left[  \bar U T^a U \right] } & \mbox{for $A=a=1,2,3$}\\
+2g_1 {\rm Tr}{\left[ \bar U U T^3 \right]} & \mbox{for $A=4$}
\end{array}
\right. .
\ee
As in appendix B, the Killing potential equations are valid by construction
\bea
{\partial \over \partial \xi^i} J_A &=& i\bar X^j_A g_{\bar j i}\nonumber\\
{\partial \over \partial \bar\xi^i} J_A &=& -i X^j_A g_{\bar i j} .
\eea
\newpage


\newpage

\begin{thebibliography}{99}

\bibitem{Nil} H. Nilles,
{\it Phys. Rep.} {\bf 110} (1984) 1.

\bibitem{Hab} H.E. Haber and G.L. Kane,
{\it Phys. Rep.} {\bf 117} (1985) 75.

\bibitem{Bar} R. Barbieri,
{\it Riv. Nuovo Cim.} {\bf 11} (1988) 1.

\bibitem{Der} J.-P. Derendinger and C.A. Savoy,
{\it Nucl. Phys.} {\bf B237} (1984) 307.

\bibitem{Drees} M. Drees,
{\it Int. J. Mod. Phys.} {\bf A4} (1989) 3635.

\bibitem{Ellis} J. Ellis, J.F. Gunion, H.E. Haber, L. Roszkowski
and F. Zwirner,
{\it Phys. Rev.} {\bf D39} (1989) 844.

\bibitem{Bin} P. Bin\'{e}truy and C.A. Savoy,
{\it Phys. Lett.} {\bf B277} (1992) 453.

\bibitem{Ellw} U. Ellwanger,
{\it Phys. Lett.} {\bf B303} (1993) 271.

\bibitem{TtV} W.T.A. ter Veldhuis,
{\it Mass Of The Lightest Higgs Boson In The Minimal Supersymmetric Standard
Model With An
Additional Singlet}, Purdue Preprint PURD-TH-92-11.

\bibitem{Elliott} T. Elliott, S.F. King and P.L. White,
{\it Phys. Lett.} {\bf B305} (1993) 71.

\bibitem{Mor} T. Moroy and Y. Okada,
{\it Phys. Lett.} {\bf B295} (1992) 73.

\bibitem{Esp} J.R. Espinosa and M. Quir\'{o}s,
{\it Phys. Lett.} {\bf B279} (1992) 92.

\bibitem{Esp2} J.R. Espinosa and M. Quir\'{o}s,
{\it Phys. Lett.} {\bf B302} (1993) 51.

\bibitem{Kane} G.L. Kane, C. Kolda and J.D. Wells,
{\it Phys. Rev. Lett.} {\bf 70} (1993) 2686.

\bibitem{Appel} T. Appelquist and C. Bernard,
{\it Phys. Rev.} {\bf D22} (1980) 200.

\bibitem{Appel2} T. Appelquist and Guo-Hong Wu,
{\it Phys. Rev.} {\bf D48} (1993) 3235.

\bibitem{Long} A. Longhitano,
{\it Phys. Rev.} {\bf D22} (1980) 1166.

\bibitem{Long2} A. Longhitano,
{\it Nucl. Phys.} {\bf B188} (1981) 118.

\bibitem{Lee} B.W. Lee, C. Quigg and H. Thacker,
{\it Phys. Rev.} {\bf D16} (1977) 1519.

\bibitem{Corn} J.M. Cornwall, D.N. Levin and G. Tiktopoulos,
{\it Phys. Rev.} {\bf D10} (1974) 1145.

\bibitem{Vay} C.E.  Vayonakis,
{\it Lett. Nuovo Cim.} {\bf 17} (1976) 383.

\bibitem{Chan} M.S. Chanowitz and M.K. Gaillard,
{\it Nucl. Phys.} {\bf B261} (1985) 379.

\bibitem{boul} G. Boulware and L.S. Brown,
{\it Ann. Phys. (N.Y.)} {\bf 138} (1982) 392.

\bibitem{Ban} M. Bando, T. Kugo and K. Yamawaki,
{\it Phys. Rep.} {\bf 164} (1988) 217.

\bibitem{Chan2} M.S. Chanowitz, M. Golden and H. Georgi,
{\it Phys. Rev.} {\bf D36} (1987) 1490.

\bibitem{Fer} S. Ferrara, A. Masiero and M. Porrati,
{\it Phys. Lett.} {\bf B301} (1993) 358.

\bibitem{Rui} R.-M. Xu,
{\it Phys. Lett.} {\bf B258} (1991) 409.

\bibitem{Buc} W. Buchm\"uller and W. Lerche,
{\it Ann. of Phys.} {\bf 175} (1987)159.

\bibitem{Khl} S. Yu. Khlebnikov,
{\it Strongly interacting W bosons and supersymmetry},
UCLA Preprint UCLA/93/TEP/29.

\bibitem{Gir} L. Girardello and M. Grisaru,
{\it Nucl. Phys.} {\bf B194} (1982) 65.

\bibitem{Cla}T. E. Clark and S. T. Love,
{\it Nucl. Phys.} {\bf B232}(1984)306,
{\it Nucl. Phys.} {\bf B254}(1985)569.

\bibitem{Bag} J. Bagger and E. Poppitz,
{\it Destabilizing Divergences in Supergravity-Coupled Supersymmetric
Theories},
Preprint JHU-TIPAC-93018.

\bibitem{Cla2} W. A. Bardeen, T. E. Clark and S. T. Love,
{\it Phys. Lett.} {\bf B237} (1990) 235;
M. Carena, T. E. Clark, C. E. M. Wagner, W. A. Bardeen and K. Sasaki,
{\it Nucl. Phys.} {\bf B369} (1992) 33.

\bibitem{Wess} J. Wess and J. Bagger,
{\it Supersymmetry and Supergravity},
Princeton University Press, 1992.

\end{thebibliography}
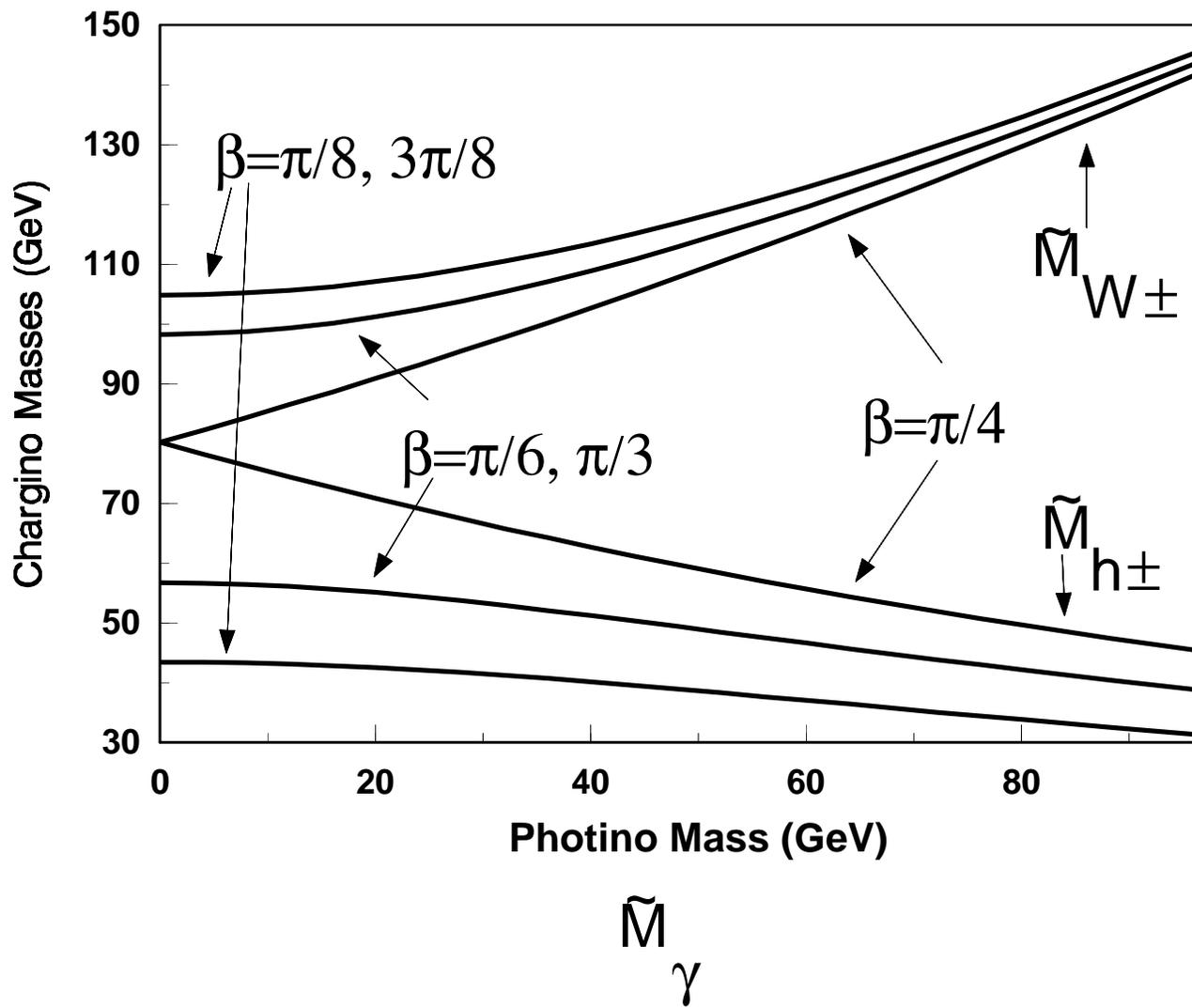
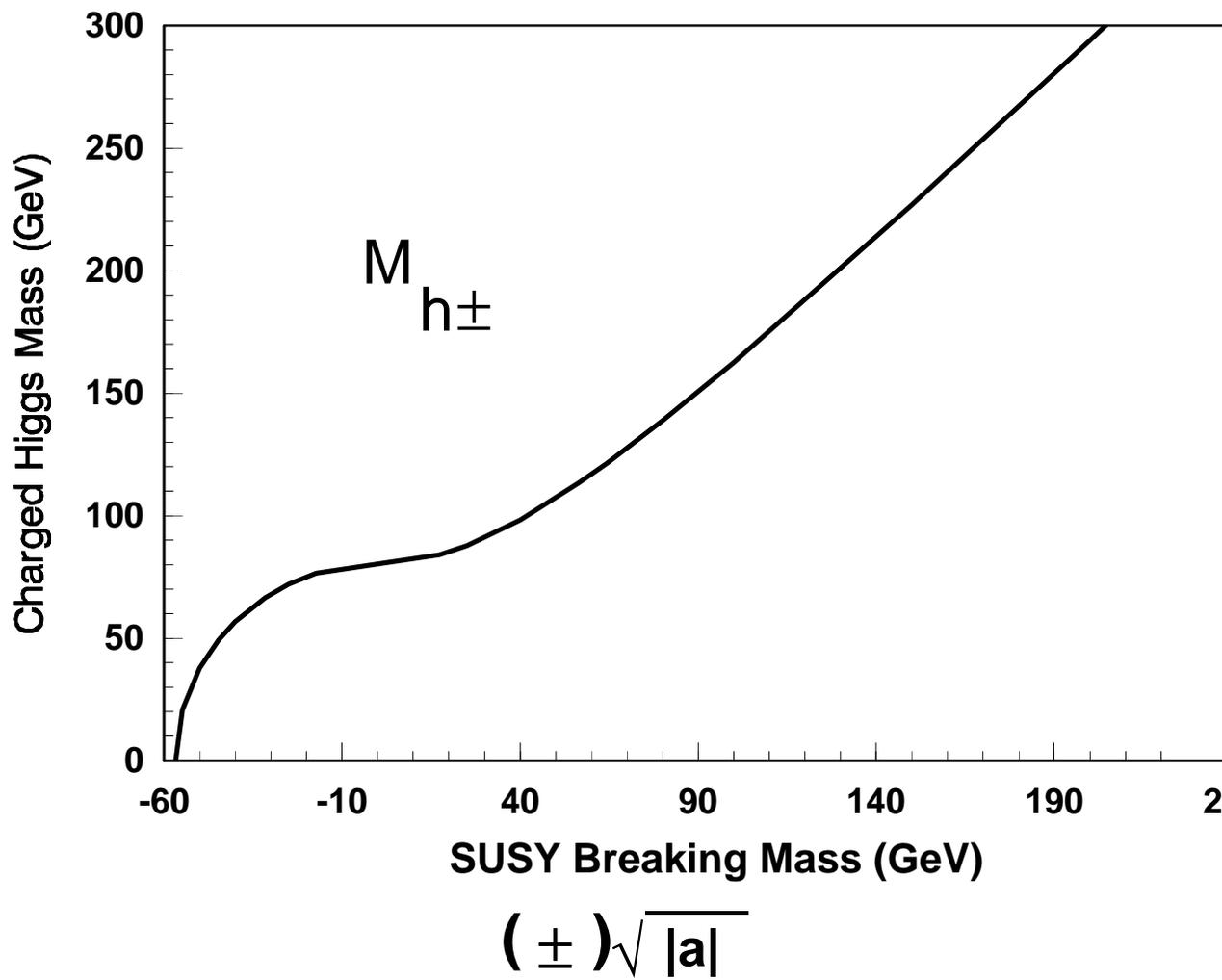
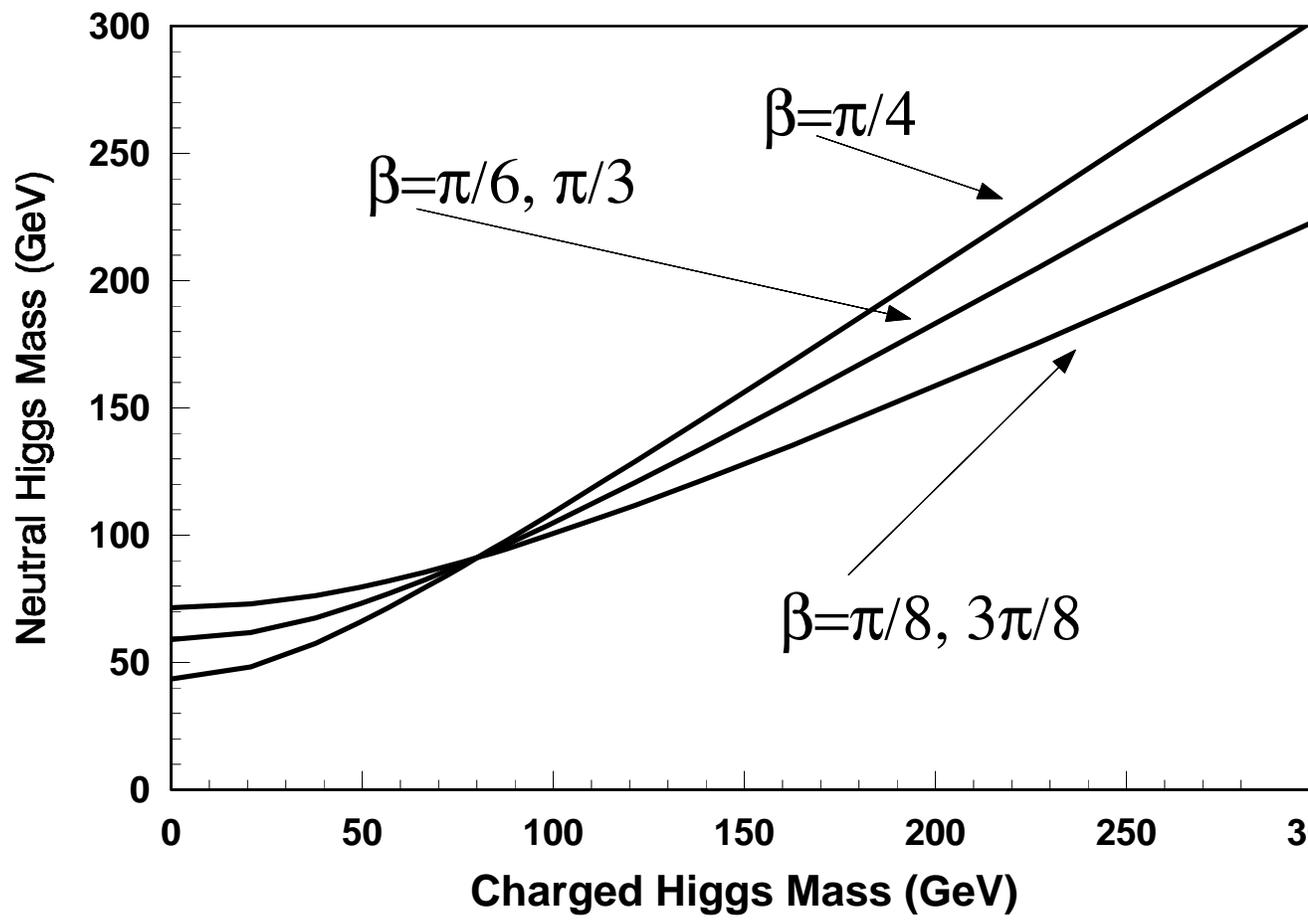
\end{document}